\documentclass[aps,12pt,final,notitlepage,oneside,onecolumn,nobibnotes,nofootinbib,superscriptaddress,showpacs,centertags]{revtex4}

\usepackage{graphicx}
\usepackage{bm}
\begin{document}

\title{ Nucleon-nucleon wave function with
short-range nodes and high-energy deuteron photodisintegration
}
\author{N.A. Khokhlov}
\email{khokhlov@fizika.khstu.ru} \affiliation{\it Pacific National
University, 680035, Khabarovsk, Russia}
\author{V.A. Knyr}
\affiliation{\it Pacific National University, 680035, Khabarovsk,
Russia}
\author{V.G. Neudatchin}
\affiliation{\it Institute of Nuclear Physics, Moscow State University, 119899, Moscow, Russia}
\date{\today}
\begin{abstract}
 We review a concept of the Moscow potential (MP) of
the $NN$ interaction. On the basis of this concept we derive by
quantum inversion optical partial potentials from the modern
partial-wave analysis (PWA) data
  and deuteron
properties. Point-form (PF) relativistic quantum mechanics (RQM) is
applied to the two-body deuteron photodisintegration. Calculations
of the cross-section angular distributions   cover photon energies
between 1.1 and 2.5 GeV. Good agreement between our theory and
recent experimental data confirms the concept of deep attractive
Moscow potential with forbidden $S$- and $P$-states.
\end{abstract}
\pacs{12.39.Pn, 13.40.-f, 13.75.Cs}
\keywords{nucleon, potential, photodisintegration, deuteron, relativistic quantum mechanics,
point-form dynamics}
\maketitle

\section{\label{sec:intro}INTRODUCTION}
Opportunities to observe manifestations of quark degrees of freedom
in nuclear reactions at intermediate energies attract attention of
scientific community for a long time. It was noted \cite{N1} that
 the most suitable subject of research here is the deuteron as
the simplest nucleus where the secondary rescattering has little
effect on the primary process.

The deuteron photodisintegration at photon energies of $\simeq 2$
GeV generates great interest among experimentalists
\cite{N1,N2,N3,N4} and theoreticians \cite{N5,N6,N7,N8,N9} with the
main emphasis on the properties of the $NN$ system which are beyond
the scope of realistic mesonic $NN$ potentials \cite{N6} and can be
interpreted within quark concepts \cite{N5,N7}. First, it was shown
in  papers by Khar'kov group \cite{N6} that starting from mesonic
potentials it is possible to explain the $d\gamma\to np$ data at
energies $E_\gamma > 1$ GeV only if a revision of electromagnetic
part of the theory is done and instead of the ordinary nucleon
electromagnetic form factors the essentially different ones are used
with poles of third order. Second, the phenomenological theory of
Regge poles was taken as a basis in Refs. \cite{N5} with selection
of dominant poles according to the quark string model \cite{N10}.
Free parameters of these theories make it possible to describe the
experimental data reasonably well. Third, also giving reasonable
results the hard rescattering model was developed \cite{N7} within a
semiempirical approach, when the photon is absorbed by a quark of
one of the nucleons and then the hard rescattering of this quark by
another nucleon takes place. The wave function amplitude of the
 final $np$-state with large  relative
 momentum is evaluated empirically by extrapolation of the
corresponding $np$-scattering experimental data.

In this paper we use the PF RQM to treat the deuteron
photodisintegration in a Poincar\'{e}-invariant way. Modern
development of the RQM and exhaustive bibliography is presented in
the review by B.D. Keister and W. Polyzou~\cite{Keister}. The PF is
one of the three forms proposed by Dirac~\cite{Dirac}. Other two are
the front form and instant form. These forms are associated with the
different possibilities for putting interactions in generators of
the Poincar\'{e} group. All the forms  are unitary equivalent
\cite{Bakamjian} but each has certain advantages. Most of the
calculations in nuclear physics have been performed in the instant
and front form. Only in recent years important simplifying features
of the PF were realized. These features are connected with the fact
that in the PF all the generators of the homogeneous Lorentz group
are free of interactions. Thus only in the PF the spectator
(impulse) approximation (SA) preserves its spectator character in
any reference frame \cite{Lev,Melde}. For an electromagnetic $NN$
process the SA implies that the $NN$ interaction  does not affect
the photon-nucleon interaction and therefore sum of the one-particle
electromagnetic current operators may be taken as an electromagnetic
current operator for the system of interacting nucleons. It is
supposed that the SA may be valid when the process is  quick due to
the large momentum transfer. General covariant PF expressions for
the electromagnetic current operator for composite systems are given
in Refs.~\cite{Lev,Klink}. The PF SA was applied to calculate form
factors of various composite particles \cite{Allen,PFSA1,PFSA2} with
reasonable results. In our calculation of the proton-proton
bremsstrahlung \cite{Myppg2} it was shown that the
 PF SA violates the continuity equation for the $NN$ current
operator, but the violation is relatively small for the considered
kinematics.

In this paper we  show that recent deuteron photodisintegration data
at $E_\gamma=1.5 - 2.5$ GeV \cite{N4} confirm the Moscow $NN$
potential model \cite{N11} characterized by deep attractive partial
potentials with forbidden $S$- and $P$-states. In this study the
Moscow partial potentials are reconstructed from the $NN$ PWA data
within the energy range $0\le E_{lab}\le 3$ GeV \cite{DataScat}.
This reconstruction is based on our approach to the
inverse-scattering problem for optical potentials~\cite{N27}.

The plan of the paper is as follows. In Sec.~II, we review a concept
of the Moscow potential (MP) of the $NN$ interaction. In Sec.~III,
we present  the optical Moscow-type $NN$ potential derived by
quantum inversion  \cite{N27} within the relativistic quasipotential
approach \cite{Keister,N20}. We show that the modern PWA data of
$NN$ scattering \cite{DataScat} are compatible with the concept of
the MP. In Sec.~IV, the formalism of PF RQM
\cite{Keister,Lev,Myppg2} is applied to the high-energy energy
deuteron photodisintegration. Results and future prospects are
discussed in Sec.~V. In Appendix A we give necessary details of the
calculation techniques. In Appendix B in the PF SA we derive an
expression for the momentum $Q_N$ transferred to the nucleon and
show that $Q_N$ is not the same as the momentum transfer seen by the
deuteron. The expression is a generalization of the similar
expression for the elastic electron-deuteron scattering
\cite{Allen}.
\section{\label{sec:IIa} POTENTIALS WITH FORBIDDEN STATES IN NUCLEAR PHYSICS}
 In description of  systems of composite $X$ particles
consisting of some $y$ particles it is a common approach to exclude
explicit degrees of freedom of $y$ particles. In simplest case of
the $XX$ system the microscopic Hamiltonian that includes all
possible pair $yy$ interactions is substituted by an effective
Hamiltonian (by sum of $X$ particle kinetic energy terms and of an
effective $XX$ potential). The common requirement is that the
effective Hamiltonian would give for the $XX$ system the same
spectrum and the same corresponding relative motion wave functions
as the initial microscopic Hamiltonian. In some cases the effective
Hamiltonian has redundant eigenvalues and eigenstates, which must be
disregarded. These eigenstates are called forbidden states and the
effective $XX$ potential is called then ''the potential with
forbidden states''.

For instance, in the oscillator shell model of the potential theory
of $\alpha-\alpha$ scattering \cite{NN24}  the antisymmetric wave
function of the $^8$Be nucleus ground state (eight-nucleon
configuration $s^4p^4$ and orbital permutation symmetry
$[f]_x=[44]$) being projected onto $\alpha-\alpha$ channel results
in $4S$-wave relative motion wave function (see our review
\cite{NN25}). This wave function accumulates all four oscillation
quanta of the system and has two nodes. Momentum distributions
corresponding to such wave functions were investigated in
quasielastic knock-out of $\alpha$ particles from $p$-shell nuclei
by  intermediate energy photons \cite{NN26}. The $0S$- and
$2S$-states of $\alpha-\alpha$ relative motion are forbidden as far
as they correspond to the lower $s^8$ and $s^6p^2$ eight-nucleon
configurations respectively, which are forbidden by the Pauli
principle. Basing on these considerations, a concept of the deep
attractive $\alpha-\alpha$ potential with $0S$, $2S$ and $2D$
forbidden bound eigenstates was elaborated \cite{NN24}. According to
the concept, there is no repulsive core in the $\alpha-\alpha$
interaction and $\alpha$ particles can penetrate into each other.
Forbidden bound eigenstates take lowest energy levels. Unforbidden
eigenstates (including scattering ones) being orthogonal to the
forbidden eigenstates have nodal structure at short range. For
instance, the $S$-wave relative motion wave function has two nodes
in the region of $\alpha-\alpha$ overlap. This model is
substantiated by the phase shift analysis based on the generalized
Levinson theorem (GLT) \cite{NN27}. For example, the $S$-wave phase
shift of $\alpha-\alpha$ scattering equals $360^{\circ}$ at zero
energy, rises up to $540^{\circ}$ at the energy slightly above the
low-lying $4S$-resonance and then runs down with increasing energy
within the broad energy range up to $E_{lab} \simeq 200$~MeV where
the phase shift approaches the asymptotic region of small values and
becomes negative due to absorption \cite{NN28}. Such picture of the
$S$-wave phase shift behavior was confirmed by experiments performed
in a broad energy range \cite{NN29}, while $D$-wave phase shift
behavior shows one forbidden state. Phase shifts of higher waves do
not show forbidden states (see \cite{NN28} for further details).

 In case of the $NN$ system, the concept of the deep attractive
$NN$ potential with forbidden states appeared in 1975 \cite{N11}
when we analyzed the $pp$-scattering phase shift data extended at
the time up to $E_{lab} \cong 6$ GeV. It was shown that   the
singlet $S$-wave phase shift data with an extended gap between low-
and high-energy groups of data  can be interpolated by a  smooth
curve if the empirical low-energy group is raised $180^{\circ}$.
This interpolation demonstrates decrease of the $S$-wave phase shift
in the
 broad energy range from zero up to $E_{lab} \gtrsim 5$ GeV
as a  manifestation of the GLT. The high-energy part
($E_{lab}\simeq3-6$~GeV) of the interpolation for the $S$-wave
remains in the asymptotic region of small values, corresponding to
the Born approximation. The energy dependence of the singlet
$D$-wave phase shift is smooth and there is no need to raise the
initial values. Calculation showed \cite{N11} that results of this
analysis are described by a deep attractive $NN$ potential with one
forbidden bound $S$-wave state. The forbidden state has a wave
function without a node. As a result, the $^{1}S_0$-wave scattering
wave function has a short-range nodal structure instead of
short-range suppression specific to a repulsive  core  potential
(RCP).  After that, a preliminary attempt was made within the
concept of MP \cite{NN30} to reconstruct $NN$ potentials for the
lowest partial waves ($S$ and $P$) from data of the $pp$ and $pn$
PWA extended at the time to intermediate energies.

At the same time the quark microscopic foundation of the MP remains
the principal problem. Unlike the nuclear shell-model picture of the
$\alpha-\alpha$ interaction the lowest quark configuration $s^6$ is
not forbidden by the Pauli principle and the corresponding $0S$-wave
state of relative $NN$ motion is not  forbidden either. Microscopic
quark investigations of the last two decades with various kinds of
$qq$ interactions have resulted in the following short-range
properties of the $NN$ system \cite{NN31}. There is a strong mixing
of different six-quark configurations in the overlap region of two
nucleons. For the $S$-wave states the leading configurations are
$s^6$ and $s^4p^2$ with comparable weights and destructive
interference. This destructive interference leads to strong
short-range suppression of the $NN$ wave function. The suppression
is described effectively by an RCP \cite{NN32}. The $s^4p^2[42]_x$
configuration introduced in our papers \cite{N11} and corresponding
to the $2S$-state of relative $NN$ motion (i.e. to the MP) would
dominate for instance in case of strong instanton induced
quark-quark interaction
  but this interaction is not strong enough \cite{Kusainov}.
Further investigations \cite{NN34,NN35} showed  that there exists a
 source for  strengthening of the $s^4p^2$ configuration. Namely, if
coupling of the $NN$,  $\Delta\Delta$ and hidden color  $CC$
channels is taken into account within the resonating group method
then the symmetry structure of the highly dominant six-quark
configuration $s^4p^2$ implies the existence of a node in the
$S$-wave relative motion wave function at short distances. Such
nodes are specific to the MP.
  In the same
  manner microscopic $qq$ interaction may give  a short-range node
  in a $P$-wave of relative $NN$ motion wave function
(in case of dominant six-quark configuration $s^3p^3[33]_x$).

In summary, the question which type of the potential (MP or RCP)
would be equivalent to the short-range quark microscopic picture of
the $NN$ interaction is highly controversial. For any RCP  a phase
equivalent supersymmetric partner with forbidden states (i.e. an MP)
may be constructed \cite{Sparenberg1}. Therefore these potentials
are indistinguishable for the $NN$ PWA. Specific to the MP
appearance of short-range nodes in  $S$- and $P$-wave relative
motion wave functions is a result of complicated six-quark dynamics
which is yet to be clarified. The nodal behavior of the MP wave
function means that the wave function is not suppressed at short
range as in case of an RCP. Thus the MP produces high-momentum
component richer than an RCP. This high-momentum component may be
seen in electromagnetic reactions with two nucleons. In
Ref.~\cite{GlozmanDeut} it was shown that the available MP produces
too rich high-momentum component in contradiction with the deuteron
electromagnetic form factors. Thus we use the latest high-energy PWA
data to refine short-range part of the MP. In our papers
\cite{Myppg1,Myppg2}  it was shown that the hard $pp\to pp\gamma$
bremsstrahlung at moderate energies ($E_{lab}\simeq 500$ MeV) is
critical to the kind of potential (MP versus RCP). The available
experimental data at smaller energy of $E_{lab} = 280$ MeV
\cite{N19} give only preliminary indication of MP validity
\cite{Myppg2}. Our present paper strengthens this line of
phenomenological research using modern deuteron photodisintegration
data.
%
%
\section{\label{sec:II}RELATIVISTIC OPTICAL $NN$ POTENTIAL}
We apply the method of inversion \cite{N27} to the analysis  of $NN$
data up to energies at which relativistic effects are essential. We
take into account these effects in the frames of the RQM
\cite{Keister,Lev}. A system of two particles is described by the
wave function, which is an eigenfunction of the mass operator
$\hat{M}$. In this case, we may represent this wave function as a
product of the external and internal wave functions. The internal
wave function $\vert\chi\rangle$ is also an eigenfunction of the
mass operator and for system of two nucleons with masses
$m_{1}=m_{2}=m$ satisfies the equation
%
\begin{equation}
\label{rel1} \hat{M}\vert\chi\rangle\equiv \left[ 2\sqrt{\hat{{\bf
q}}^2+m^2}+V_{int} \right]\vert\chi\rangle = M\vert\chi\rangle,
\end{equation}
where $V_{int}$ is  an operator commuting with the full angular
momentum operator and acting only through internal variables (spins
and relative momentum), $\hat{{\bf q}}$ is a momentum operator of
one of the particles in the center of mass frame (relative
momentum). Rearrangement of (\ref{rel1}) gives
\begin{equation}
\label{rel2} \left[ {\hat{{\bf q}}^2 + m V} \right]\chi = q^2\chi,
\end{equation}
where $V$ acts like $V_{int}$ only through internal variables and
\begin{equation}
  q^2 = \frac{M^2}{4} - 2m^2.\label{q2}
\end{equation}
Eq.~(\ref{rel2}) is identical in form to the Schr\^{o}dinger
equation. The formally same equation may be deduced as a truncation
of the quantum field dynamics \cite{N20}. The quasicoordinate
representation corresponds to the realization ${\hat {\bf q}} = -
i\frac{\partial }{\partial {\rm {\bf r}}}$, $V=V({\rm {\bf r}})$.

We applied the method of inversion \cite{N27}   to reconstruction of
the nucleon-nucleon partial potentials
\begin{equation}
V(r)=(1+i\alpha)V^{(0)}(r),
\end{equation}
for single waves and
\begin{equation}
\label{eq28} V\left( r \right) =\left( {{\begin{array}{*{20}c}
 {\left( {1 + i\alpha_{1} } \right)V_{1}^{(0)}(r) } \hfill & {\left( 1 + i
\alpha_{3}
\right)V_{T}^{(0)} (r)} \hfill \\
{\left( 1 + i \alpha_{3} \right)V_{T}^{(0)} (r)} & {\left( {1 +
i\alpha_{2} } \right)V_{2}^{(0)}
(r)} \hfill \\
\end{array} }} \right),
\end{equation}
for coupled waves, where $V^{(0)}(r)$ are energy-independent real
and inelasticity parameters $\alpha$
depend on energy. 
As input data for the reconstruction
 we used  modern PWA data (single-energy solutions)
 up to  $1200$ MeV for isoscalar states and up to $3$ GeV for isovector states of the $NN$
 system \cite{DataScat}. The deuteron properties
were taken from \cite{DataDeut}. These data allow us to construct
Moscow-type $NN$ partial potentials sustaining forbidden bound
states. These potentials describe part of the deuteron properties
and the PWA data by the construction.  

According to the MP concept and the GLT  some phase shift data of
\cite{DataScat} are raised $180^{\circ}$.  Namely, $^{1}S_{0}$-wave
phase shift and all four $^{2S+1}P_J$-wave phase shifts are equal to
$180^{\circ}$ at zero energy, $^{3}S_{1}$-wave phase shift is equal
to $360^{\circ}$ at zero energy. The mixing parameters $\epsilon_1$
and $\epsilon_2$ of the MP differ from that of a traditional RCP by
sign. All phase shifts for higher waves (for $L\ge 2$) are
''small'', they have zero values at zero energy. According to our
model  we have fitted free parameters of the inversion solutions  to
get nodes at $r \simeq$ 0.5 fm in $S$ and $P$ waves and to make
central parts of the potentials close to each other and to the
Gaussian shape. The energies of forbidden states are in the range
$300-750$ MeV.

\begin{figure}[htbp!]
\includegraphics[width=8.6cm]{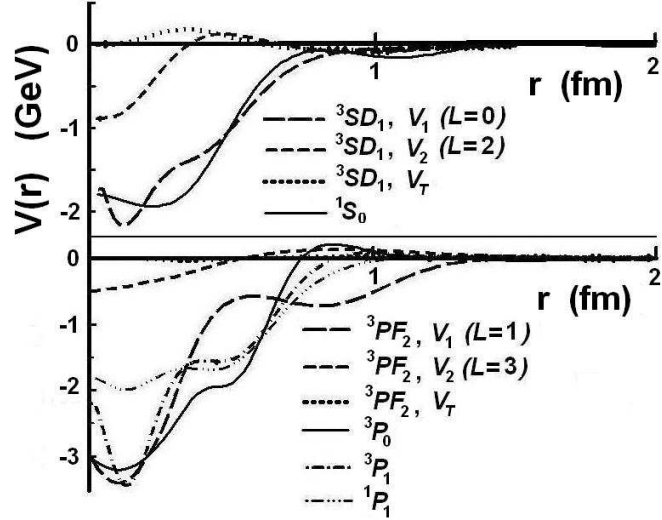}
\caption{\label{fig1} Reconstructed partial potentials  for  lower
orbital momentum (single and coupled channels).}
\end{figure}

\begin{figure}[htbp!]
\includegraphics[width=8.6cm]{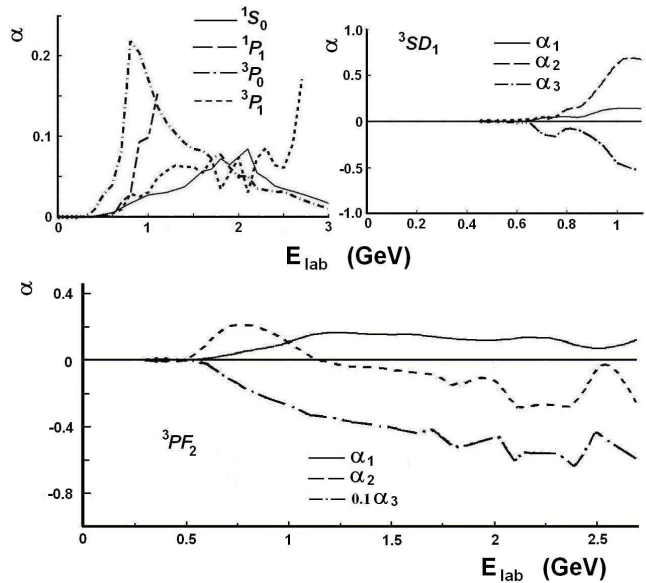}
\caption{\label{fig2} Reconstructed  inelasticity multipliers
$\alpha$ for the  potentials presented in Fig.~\ref{fig1}. }
\end{figure}

\begin{figure}[htbp!]
\includegraphics[width=15.0cm]{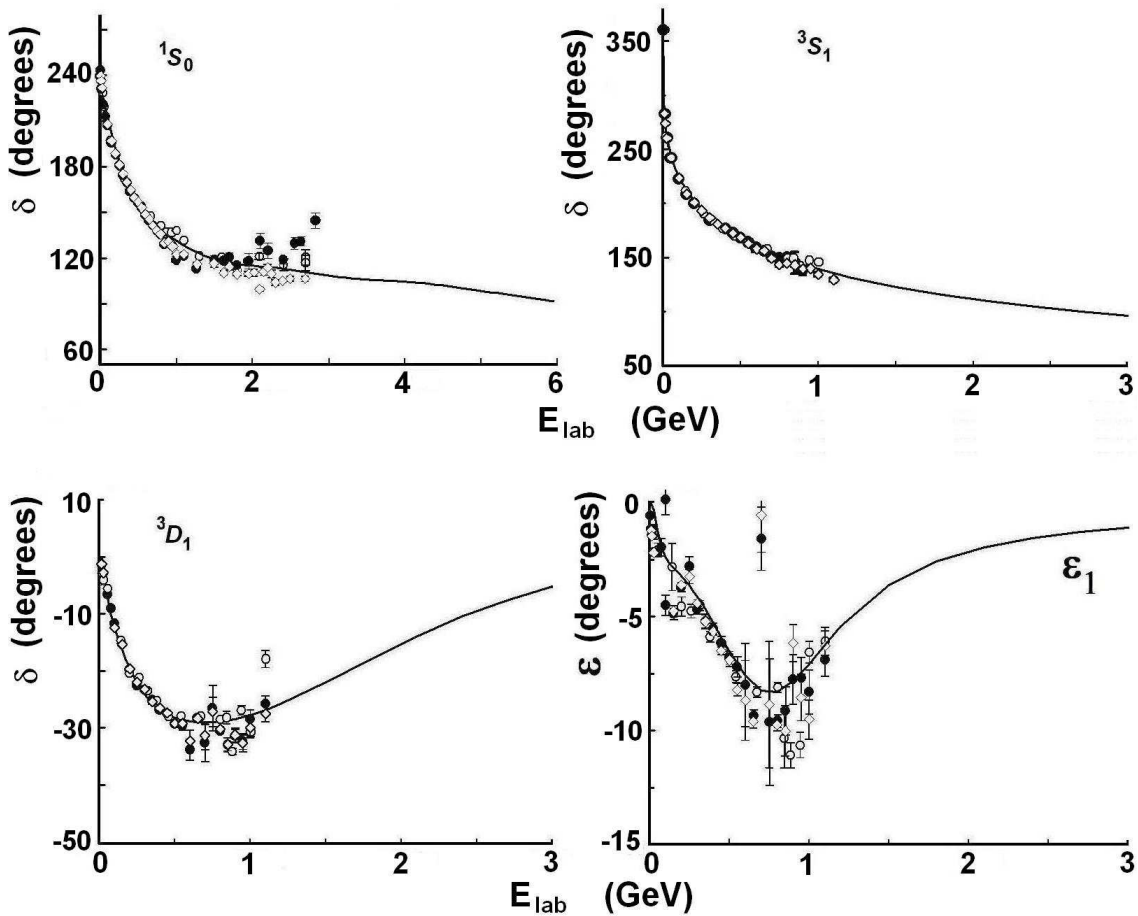}
\caption{\label{fig3} Phase shifts and mixing parameters in the
present optical model. PWA data are from \cite{DataScat}. For
$^{1}S_0$ and $^{2S+1}P_J$ waves, the original data set  from
Ref.~\cite{DataScat} is raised $180^{\circ}$. To leave the $S$
matrix unchanged, we change the sign of the mixing parameters
$\varepsilon_1$ and $\varepsilon_2$. }
\end{figure}

Our calculations show that the final state interaction (FSI) in the
$S$ and $P$ waves gives by far the largest contribution to the
deuteron photodisintegration cross-section comparing with FSI in
other waves, so we present  results of inversion only for these and
for coupled to them waves.
 Part of results presented in Figs. \ref{fig1}-\ref{fig4} (for $^1 S_0$ and $^3
SD_1$ waves) we presented earlier in Ref.~\cite{N27}.

The reconstructed potentials $V^{(0)}(r)$  are displayed in
Fig.~\ref{fig1}. The inelasticity multipliers $\alpha$ are displayed
in Fig.~\ref{fig2}. Fig.~\ref{fig3} displays reproduction of the
corresponding  phase shifts and mixing parameters. In
Fig.~\ref{fig4} the description of inelasticity parameters is shown.
All the $P$-wave phase shifts are positive according to the GLT.
Large difference between $^{3}P_{0}$-wave and $^{3}P_{2}$-wave phase
shift curves reflects a large spin-orbital interaction which is
attractive for the $^{3}P_{2}$-wave as we see. These features
correspond to the general properties of the MP (its large positive
gradient in the region $r<1$ fm). It is interesting to learn from
Fig. \ref{fig3} that among four lowest $pp$ phase shifts  three of
them ($^{1}S_{0}$, $^{3}P_{0}$ and $^{3}P_{2}$) correspond to the MP
but the experimental data within the energy range $E_{lab}=2-3$~GeV
are contradictory  for the $^{3}P_{1}$-wave phase shift. It would be
important to refine the PWA data in this range using modern
polarization data on $pp$-scattering.  The $S$- and $D$-state wave
functions of deuteron are displayed in Fig. \ref{fig5}. There is a
node in the $S$-wave function at $r\simeq 0.5$ fm and both wave
functions are not suppressed at short-range in contrast with wave
functions produced by an RCP. For continuum $S$- and $P$-wave
functions the node radii equal to $0.5-0.9$ fm at the considered
energies. All potentials and inelasticity multiplies ($\alpha$'s)
can be accessed via a link to the website \cite{page}.
\setcounter{figure}{2}
\begin{figure}[htbp!]
\includegraphics[width=15.cm]{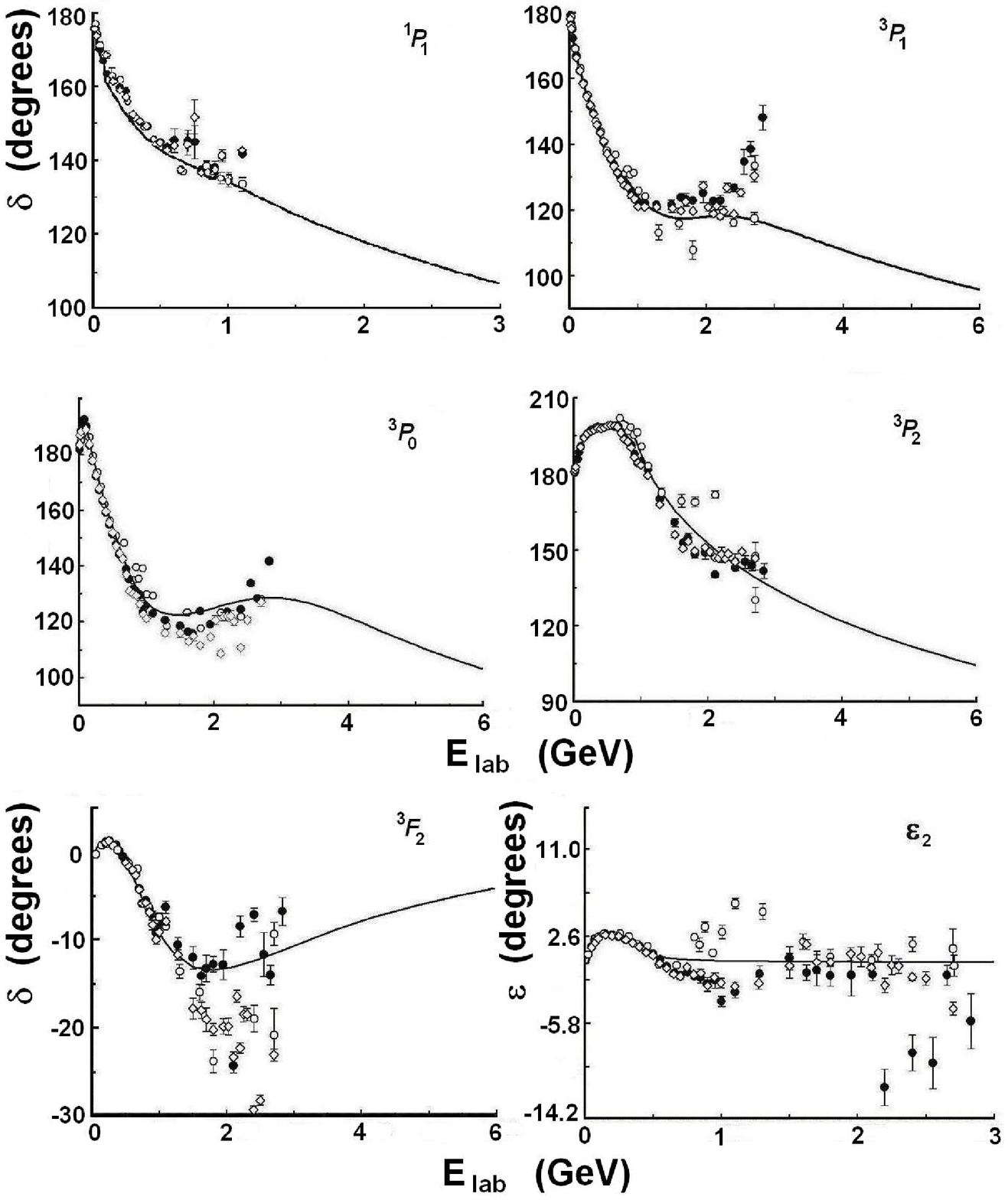}
\caption{\label{fig3b} (Continued). }
\end{figure}
\begin{figure}[htbp!]
\includegraphics[width=15.cm]{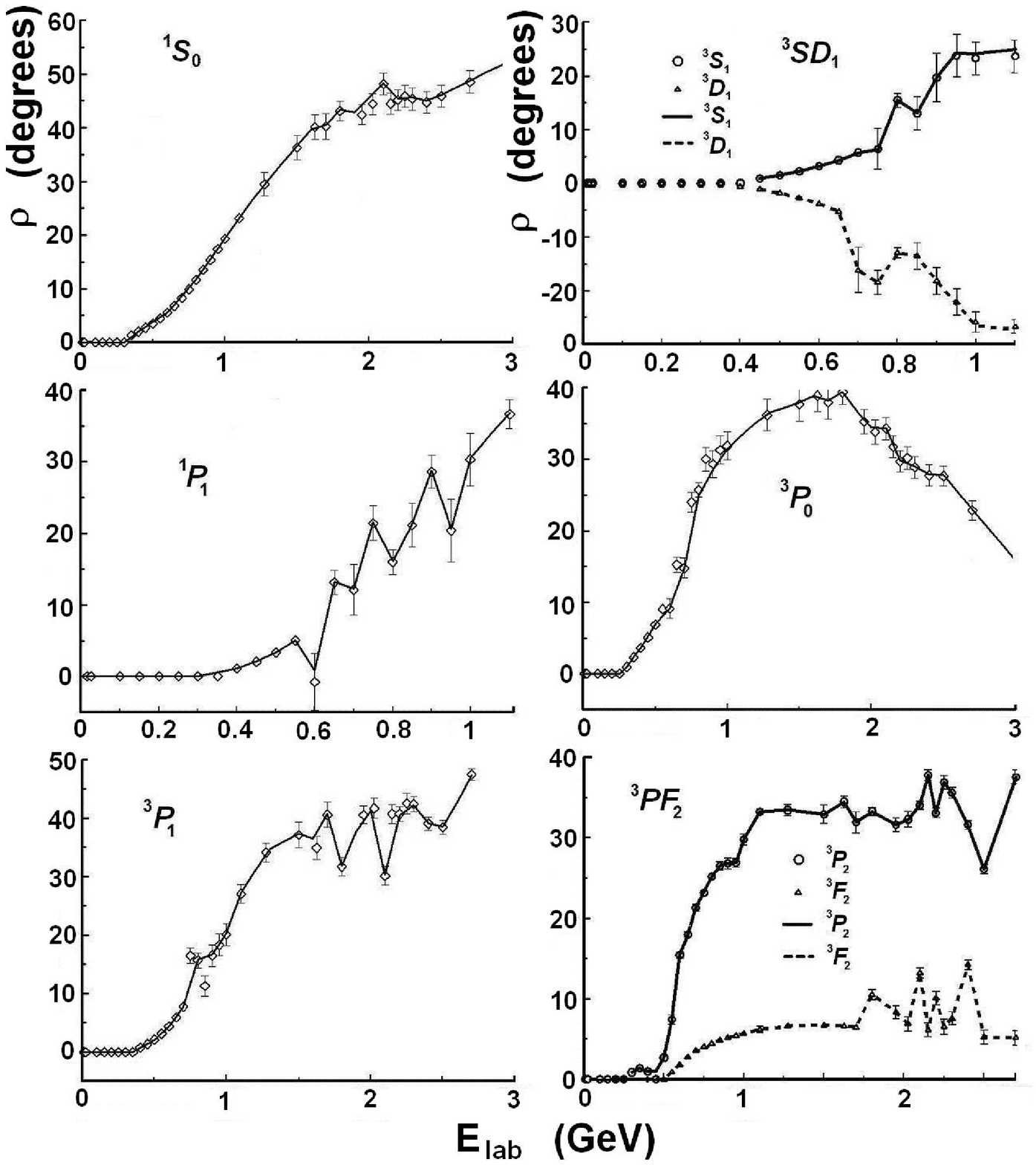}
\caption{\label{fig4} Inelasticity parameters $\rho$  in the present
optical model. PWA data are from \cite{DataScat}.}
\end{figure}
\begin{figure}[htbp!]
\includegraphics[width=8.6cm]{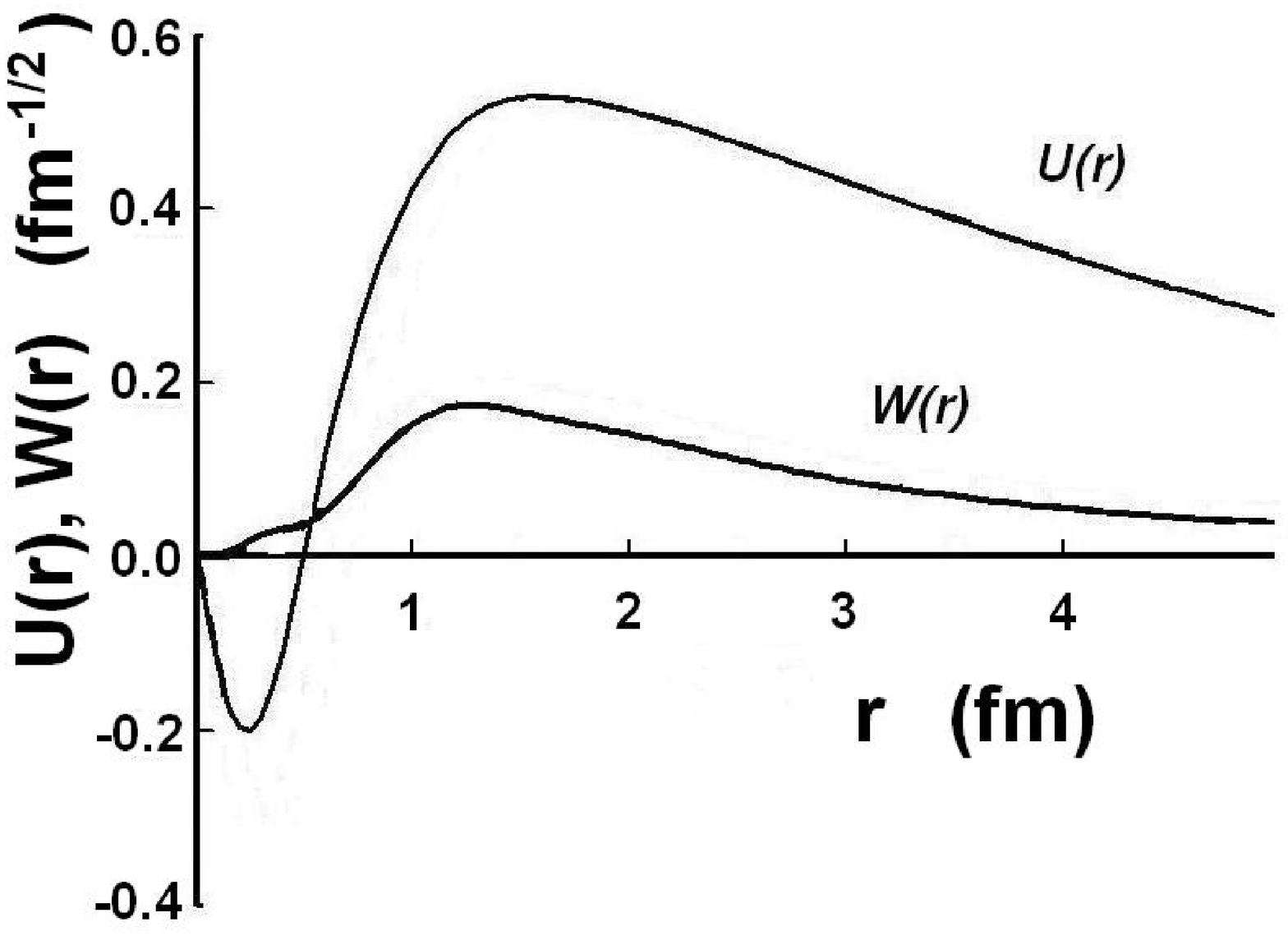}
\caption{\label{fig5} The deuteron $S$-wave and $D$-wave functions
for present version of the $NN$ Moscow potential.}
\end{figure}

It should be pointed out that in nuclear matter calculations the
$NN$ potentials  should be used in the following form
\begin{equation}
V^{nucl}(r)=V^{(0)}(r)+\lambda \langle \chi_{S,L,J}\vert\,
\end{equation}
where operator $\langle \chi_{S,L,J}\vert$ projects onto the
forbidden state $\vert \chi_{S,L,J}\rangle$, positive constant
$\lambda$ tends to infinity. The forbidden state $\vert
\chi_{S,L,J}\rangle$ may be found from Eq.~(\ref{rel2}) by some
numerical method as a bound state of the partial potential
$V^{(0)}(r)$ (all bound states are forbidden except the deuteron
one). Constant $\lambda$ is a large number, such that its further
increase does not change results of calculation. This procedure
orthogonalizes the nuclear wave function to forbidden two-nucleon
states. Thus, we exclude the unphysical collapse of nuclear matter.
\section{\label{sec:III}DEUTERON
PHOTODISINTEGRATION IN POINT-FORM RELATIVISTIC QUANTUM MECHANICS}
 Formalism of the PF is considered in detail in Refs.~\cite{Keister,Lev}, while
  general covariant PF expressions for the
electromagnetic current operator for composite systems  are given in
Refs.~\cite{Lev,Klink}.  Therefore, we give only results necessary
for our calculation, in notation of Ref.~\cite{Lev}. We use
algorithm of \cite{Lev} for calculation of the matrix elements of
the electromagnetic current operator. We applied this formalism to
the $pp\gamma$ process \cite{Myppg2}. Similar approach was applied
to the elastic electron-deuteron scattering \cite{Allen}.

We consider the $pn$ system and neglect difference of neutron and
proton masses ($m_1=m_2=m$). Let $p_i$ be the 4-momentum  of nucleon
$i$, $P\equiv(P^0,{\bf P})=p_1+p_2$ be the system 4-momentum, $M$ be
the system mass and $G=P/M$ be the system 4-velocity. The wave
function of two particles with 4-momentum $P$ is expressed through a
tensor product of external and internal parts
\begin{equation}
\vert P,\chi\rangle
=U_{12}\,\vert P\rangle\otimes\vert \chi\rangle,
\end{equation}
where the internal wave function $\vert\chi\rangle$ satisfies
Eqs.~(\ref{rel1})-(\ref{rel2}). The operator
\begin{equation}
U_{12}=U_{12}(G,{\bf q})=\prod_{i=1}^{2} D[{\bf
s}_i;\alpha(p_i/m)^{-1}\alpha(G)\alpha(q_i/m)] \label{U12}
\end{equation}
is the unitary operator from the ''internal'' Hilbert space  to the
Hilbert representation space of two-particle states \cite{Lev}.
$D[{\bf s};u]$ is the representation operator of the group SU(2)
corresponding to the element $u\in$SU(2) for the representation with
the generators ${\bf s}$. Action of  $D[{\bf s};u]$ and matrices
$\alpha$ are defined in Appendix A, $s_i=1/2$ is spin of a nucleon.
The momenta of the particles in their c.m. frame are
\begin{equation}
q_i=L[\alpha(G)]^{-1}p_i, \label{scm_q}
\end{equation}
where $L[\alpha(G)]$ is the Lorentz transformation to the frame
moving with 4-velocity $G$ ($L[\alpha(G)]^{-1}$ is the inverse
transformation). It is easy to verify that ${\bf q}_1=-{\bf q}_2$.

The external part of the wave function is defined as
\begin{eqnarray}
\label{ext_wf}  \langle G\vert
P'\rangle\equiv\frac{2}{M'}G^{'0}\delta^3({\bf G}-{\bf G}'),
\end{eqnarray}
with scalar product
\begin{eqnarray}
\label{ext_wf_product}  \langle P''\vert P'\rangle=\int
\frac{d^3{\bf G}}{2G^0} \langle P''\vert G\rangle \langle G\vert
P'\rangle=2\sqrt{M^{'2}+{\bf P}^{'2}}\delta^3({\bf P''}-{\bf P}'),
\end{eqnarray}
The internal part of the wave function
 $\vert\chi\rangle$ is characterized by momentum  ${\bf q}={\bf q}_1=-{\bf q}_2$ of one of the particles in the
 c.m. frame.
%
%
%
%
%
 Interaction appears
according to the Bakamjian---Thomas procedure
$\hat{P}=\hat{G}\hat{M}$, where $\hat{M}$ is sum of the free mass
operator ${M}$ and of the interaction $V$: $\hat{M}={M}+V_{int}$
(compare with Eq.~(\ref{rel1})). The interaction operator acts only
through internal variables. Operators $\hat{M}$, ${M}$, $V_{int}$
and $V$ commute with spin operator $S$ (full angular momentum) and
with 4-velocity operator $\hat{G}$. Interaction term is present in
all components of total 4-momentum. Generators of Lorentz boosts and
generators of rotations are free of interaction. In the c.m. frame
the relative orbital angular momentum and spins are
 coupled together as in the non-relativistic case. Moreover
most non-relativistic scattering theory  formal results are valid
for our case of two particles \cite{Keister}.

The deuteron wave function $\vert P_{i},\chi_{i}\rangle$ is
normalized as follows
\begin{equation}
\label{sc_prod_deut} \langle P_i',\chi_{i}\vert
P_i'',\chi_{i}\rangle=2 P_{i}^{0'}\,\delta^3({\bf P}_i'-{\bf
P}_i'').
\end{equation}
For one-particle wave functions  normalized in the same manner the
free two-particle states are normalized as
\begin{eqnarray}
\langle P',\chi'\vert P'',\chi''\rangle\equiv\langle p_{1}'\vert
p_{1}''\rangle \langle p_{2}'\vert p_{2}''\rangle \delta_{\mu_1'
\mu_1''}\delta_{\mu_2' \mu_2''}=
%
%
\nonumber\\
= 4w({\bf p}_1')w({\bf p}_2')\delta^{3}({\bf p}_{1}''-{\bf
p}_{1}')\delta^{3}({\bf p}_{2}''-{\bf p}_{2}')\delta_{\mu_1'
\mu_1''}\delta_{\mu_2' \mu_2''}=
\nonumber\\= 2W({\bf P}')\delta^{3}({\bf P}''-{\bf
P}')\frac{2w^2({\bf q})}{M({\bf q})}\delta^{3}({\bf q}''-{\bf
q}')\delta_{\mu_1' \mu_1''}\delta_{\mu_2' \mu_2''}=
 \nonumber\\=
 2W({\bf P}')\delta^{3}({\bf P}''-{\bf
P}')\frac{M({\bf q})}{2}\delta^{3}({\bf q}''-{\bf q}')\delta_{\mu_1'
\mu_1''}\delta_{\mu_2' \mu_2''}
\label{free two state norm}
\end{eqnarray}
where $w({\bf p})\equiv\sqrt{m^2+{\bf p}^2}$, $M({\bf q})\equiv
2\sqrt{m^2+{\bf q}^2}$, $W({\bf P})\equiv\sqrt{M^2+{\bf P}^2}$,
$G_{0}({\bf G})\equiv\sqrt{1+{\bf G}^2}$, $\mu_i$ are spin
projections in the c.m. frame. Multiplier $\frac{M({\bf q})}{2}$ is
a relativistic invariant, therefore we may normalize internal part
of the scattering state wave function in the non-relativistic manner
\begin{eqnarray}
\langle P',\chi'\vert P'',\chi''\rangle_{n.r.}=
 \nonumber\\=
 2W({\bf P}')\delta^{3}({\bf P}''-{\bf
P}')\delta^{3}({\bf q}''-{\bf q}')\delta_{S' S''}\delta_{\mu'
\mu''},
\label{free two state norm_nr}
\end{eqnarray}
 where $S$ and $\mu$ are full angular momentum and its projection in the c.m. frame.

%

The  differential cross section  for the $\gamma d\rightarrow np$
process is given by
\begin{equation}
\frac{d\sigma}{d\Omega}=\frac{q_{f}}{64\pi^{2}M_f^2k_{c}}|{A}_{if}|^{2},\label{cross2}
\end{equation}
where $q_f$ is the final asymptotic $np$ relative momentum, $k_{c}$
is photon energy in c.m. frame.
 The $d\gamma\to np$ amplitude $A_{if}$ is defined in the same
manner as the $pp\gamma$ amplitude that was used in
Ref.~\cite{Myppg2}
\begin{eqnarray}
 (2\pi)^4\,\delta^4(P_i+k-P_f)\,A_{if}= \nonumber\\=\,
\sqrt{4\pi}\int d^4 x\, \langle
P_f,\chi_f\vert\,\varepsilon_{\mu}\hat{J}^{\mu}(x)\,\vert
P_i,\chi_{i}\rangle\,e^{i kx}\label{amplit_def}
\end{eqnarray}
where $P_i$ and $P_f$ are initial and final 4-momenta of the $NN$
system correspondingly, $\varepsilon_{\mu}$ is the photon
polarization vector.

Following Ref.~\cite{Lev} we choose for calculation of the invariant
amplitude ${A}_{if}$ a special frame defined by condition
\begin{equation}
\label{LevFrame} {\bf G}_{i}+{\bf G}_f=0,
\end{equation}
where ${G}_i={P}_i/M_i$, ${G}_f={P}_f/M_f$ are  4-velocities of
initial and final $NN$ c.m. frames respectively (${\bf G}_{i}$ and
${\bf G}_f$ are their 3-vector parts). The initial mass $M_i$ is the
deuteron mass. The final mass $M_f$ is the invariant mass of the
final $NN$ system.  These masses are different due to absorption of
a photon, therefore coordinate frame corresponding to Eq.
(\ref{LevFrame}) is not equivalent to the Breit frame where ${\bf
P}_i+{\bf P}_f=0$. Masses $M_i$ and $M_f$ define also corresponding
wave functions through Eq.~(\ref{q2}) and Eq.~(\ref{rel2}).

The matrix elements of the current operator $\hat{J}^{\mu}(x)$
appears especially simple
 in the frame defined by Eq.~(\ref{LevFrame}):
\begin{eqnarray}
 \langle P_f,\chi_f\vert\,\hat{J}^{\mu}(x)\,\vert
P_i,\chi_i\rangle=\nonumber\\=\,4\pi^{3/2}\sqrt{M_i} M_f\, e^{i
(P_f-P_i)x}\,\langle \chi_f\vert\,\hat{j}^{\mu}({\bf h})\,\vert
\chi_i\rangle_{n.r.}, \label{MA_of_CO}
\end{eqnarray}
where $\hat{j}^{\mu}({\bf h})$ is the current operator in the
internal space defined in Ref.~\cite{Lev} as an operator ${\hat
J}(0)$ (see Eq.~(\ref{J0})) in the frame (\ref{LevFrame}) expressed
through ${\bf h}$ and ${\bf q}$. Following Ref.~\cite{Lev} we use
the dimensionless vector ${\bf h}={\bf G}_f/G_f^0$, where $G_f$ is a
4-velocity of the final $NN$ system in the frame defined by
Eq.~(\ref{LevFrame}).
This parameter may be expressed through the photon momentum ${\bf k}$, so that 
 ${\bf  h}=2(M_i
M_f)^{1/2}(M_i+M_f)^{-2}\,{\bf  k}$, $\vert{\bf  h}\vert\equiv
\,h=(M_i-M_f)/(M_i+M_f)<1$. Convenience of this parameter is
illustrated in Appendix B.

 The internal wave functions of the deuteron  and
of the final scattering state are normalized in the non-relativistic
manner.
 The deuteron wave
function is
\begin{equation}
\label{eq_12} \vert
\chi_i\rangle=\vert\chi_i\rangle_{n.r.}=\frac{1}{r}\,\sum_{l=0,2}\,u_l(r)\,\vert
l,1;\,1 M_J\rangle\, ,
\end{equation}
with normalization $\langle\chi_i\vert\chi_i\rangle_{n.r.}=1$, where
\begin{equation}
 \qquad \vert l,S;\,J M_J\rangle=\sum_{m}\sum_{\mu}\,\vert
S,\mu\rangle\, {\cal Y}_{lm}(\hat{n})\,{\cal C}^{J M_{J}}_{l m S
\mu}\, .
\end{equation}
 The
  internal  wave function of the final continuum $np$ state  is
\begin{widetext}
\begin{eqnarray}\nonumber
\vert\chi_f\rangle\equiv\vert q_{f},S_f,\mu_f\rangle_{n.r.} =
\sqrt{\frac{2}{\pi}}\frac{1}{q_{f}r}\sum_{J=0}^{\infty}\times\,\\
\times\sum_{M_J=-J}^{J}\,\sum_{l=J-S}^{J+S}\,\sum_{l'=J-S}^{J+S}\,\sum_{m=-l}^{l}
i^{l'}u^{J}_{l',l}(q_{f},r){\cal C}^{JM_{J}}_{lm\,S_f\mu_f}{\cal
Y}^{\ast}_{lm}(\hat{q}_{f})\vert l',S_f;JM_{J} \rangle,
\label{wf_scat}
\end{eqnarray}
\end{widetext}
with normalization
$\langle\chi_{f'}\vert\chi_{f}\rangle_{n.r.}=\delta({\bf
q}_f^{\,\prime}-{\bf q}_f)\delta_{S_f\,S_f'}\delta_{\mu_f\,\mu_f'}$.
The corresponding plane wave $\vert\phi_f\rangle_{n.r.}$ is
characterized by the spherical Bessel functions
$j_l(q_f,r)\delta_{ll'}$
 instead of $u^{J}_{l,l^{\prime}}(q_{f},r)$. The deuteron partial wave
 functions $u_l(r)$ presented in Fig.~\ref{fig5} and partial waves
 of the final $np$ states $u^{J}_{l,l^{\prime}}(q_{f},r)$ are
 calculated from Eq.~(\ref{rel2}).

 We define a reduced amplitude
 \begin{equation}
T_{fi}=\langle
\chi_f\vert\,\varepsilon^{\ast}_{\mu}\hat{j}^{\mu}({\bf  h})\,\vert
\chi_i\rangle_{n.r.}.\label{reducedT}
\end{equation}%
 As a result the differential
cross-section (\ref{cross2}) can be rewritten as
\begin{equation}
\label{eq_10} \frac{d\sigma}{d\Omega}=\frac{\pi^2
q_{f}M_{i}}{6k_{c}}\sum_{i}\sum_{f}|T_{fi}|^{2}\,.
\end{equation}
where we average over photon polarizations,
 spin orientations of initial deuteron and sum over spin orientations of final nucleons.

In our calculations we  approximate the above matrix element
\begin{equation}
\label{matixElAll} \langle \chi_f\vert\,\hat{j}^{\mu}({\bf
h})\,\vert \chi_i\rangle_{n.r.} \approx \langle
\phi_f\vert\,\hat{j}^{\mu}({\bf h})\,\vert
\chi_i\rangle_{n.r.}+\langle
\chi_f-\phi_f\vert\,\hat{\tilde{j}}^{\mu}({\bf  h})\,\vert
\chi_i\rangle_{n.r.}\, .
\end{equation}
The first term is a plane wave approximation (PlWA) and is
calculated using the exact current operator (\ref{exact_current}).
In this case the operator $\hat{{\bf q}}$ can be substituted by
${\bf q}_f$ and operator structure of $\hat{j}^{\mu}({\bf  h})$ can
be presented as
\begin{eqnarray}
 \hat{j}^{\mu}( {\bf  h})={j}^{\mu}( {\bf  h})+\delta{j^{\mu}}=\sum_{i=1,2}\left( {
B}_{1i}^{\mu}+({\bf  B}^{\mu}_{2i}\cdot{\bf  s}_i)+({\bf
B}^{\mu}_{3i}\cdot{\bf  s}_k )+\right.\nonumber\\\left.+\,({\bf
B}^{\mu}_{4i}\cdot{\bf  s}_i)({\bf  B}^{\mu}_{5i}\cdot{\bf  s}_k
)\right){ I}_{i}({\bf  h})+\delta{  j}\, , \label{preciseCurr}
\end{eqnarray}
where ${j}^{\mu}( {\bf  h})$ is sum of the one-nucleon
electromagnetic current operators (spectator approximation), addend
$\delta {j}^{\mu}$ restores the current conservation equation;
 $k$ = 2, if $i$ =1, and,
conversely, $k$ = 1, if $i$ = 2. ${ B}^{\mu}_{1i}$ and ${\bf
B}^{\mu}_{mi}\, ,\,\, m\ge 2$ are  vector and tensor functions of
arguments ${\bf  h}$ and ${\bf  q}_f$. These functions are given in
Appendix A. In Appendix A we calculate addend $\delta { j}^{\mu}$
from the current conservation equation following Ref.~\cite{Lev} as
we did for the $pp\gamma$ process in Ref.~\cite{Myppg2}. Obviously
our phenomenological quasipotential model offers no microscopic
picture of interaction that would allow us to  unambiguously
determine the current operator. We use the defined bellow  $\delta {
j}^{\mu}$ only to estimate violation of the current conservation
equation. Assuming gauge invariance (which follows from the
Poincar\'{e} invariance and the current conservation equation) we
use the transverse gauge
\begin{equation}
\varepsilon_{\mu}=(0,{\bf \varepsilon}),\ \ \ \ \ ({\bf
\varepsilon}{\bf k})=0.\label{trgauge}
\end{equation}
Thus, we exclude the ${j}^{0}( {\bf  h})$ and ${j}_{||}( {\bf h})$
(see Eq.~(\ref{decomposition})) components of the current
from Eq.~(\ref{reducedT}). %
The Poincar\'{e} invariance is ensured by definition of the current
operator $\hat{J}^{\mu}(x)$ through the operator $\hat{j}^{\mu}({\bf
h})$ (see details in Ref.~\cite{Lev}).

 The use of $(\chi({\bf  r})-\phi({\bf  r}))$ combination in Eq.~(\ref{matixElAll})
accelerates convergence of the partial wave expansion. This term is
nonzero due to FSI of neutron and proton. It is calculated from the
first order in $h$ approximation of the current operator
$\hat{j}^{\mu}( {\bf  h})$.  This approximation calculated in the
same manner as one for the $pp$ system in Ref. \cite{Myppg2} is
given by
\begin{widetext}
\begin{eqnarray}
 \nonumber
\hat{{{\bf j}}}({\bf h})\approx\hat{\tilde{{\bf j}}}({\bf
h})=\delta{\bf  j}+ \frac{{\bf  q}}{w}\hat{g}^{pn}_{e}(0) - {\bf
h}\hat{G}^{pn}_{e}(0)+\nonumber\\
 + \imath \left(\frac{m}{w}\lbrack
{\bf  S}\times {\bf  h}\rbrack + \frac{1}{w(w+m)} \lbrack {\bf
q}\times {\bf  h}\rbrack\left({\bf  q}\cdot {\bf
S}\right)\right)\hat{G}_{m}^{pn}(0)
 +\nonumber\\+\imath \left(\frac{m}{w}\lbrack {\bf  T}\times {\bf  h}\rbrack +
\frac{1}{w(w+m)} \lbrack {\bf  q}\times {\bf  h}\rbrack\left({\bf
q}\cdot {\bf  T}\right)\right)\hat{g}_{m}^{pn}(0) +\nonumber\\\imath
\,({\bf h}\cdot \lbrack {\bf  q}\times{\bf S}\rbrack )\,{\bf
q}\left(\frac{\hat{G}_{m}^{pn}(0)}{mw}+\frac{\hat{G}_{e}^{pn}(0)}{w(w+m)}\right)+
\nonumber\\
+\imath ({\bf  h}\cdot \lbrack {\bf  q}\times{\bf  T}\rbrack )\,{\bf
q}\left(\frac{\hat{g}_{m}^{pn}(0)}{mw}+\frac{\hat{g}_{e}^{pn}(0)}{w(w+m)}\right)-({\bf
h}\cdot {\bf  q})\,{\bf  q}\,\frac{\hat{G}^{pn}_{e}(0)}{mw}
\,,\qquad\qquad\label{FSIcurrent}
\end{eqnarray}
\begin{eqnarray}
\delta{\bf  j}=\left(\frac{4w}{M_f+M_i}-1-h\right) \frac{{\bf
q}}{w}\hat{g}_{e}^{pn}(0)+ \nonumber\\+
 \imath h\left( \lbrack {\bf  q}\times {\bf
T}\rbrack\left(\frac{\hat{G}_{m}^{pn}(0)}{m}
-\frac{\hat{G}_{e}^{pn}(0)}{w+m}\right)
 -2\hat{g}^{pn}_{e}(0)\,w\,{\bf  r}+\lbrack {\bf  q}\times {\bf
S}\rbrack\left(\frac{\hat{g}_{m}^{pn}(0)}{m}-\frac{\hat{g}^{pn}_{e}(0)}{w+m}\right)\right),
\label{dj}
\end{eqnarray}
\end{widetext}
where ${\bf  S}={\bf  s}_1+{\bf  s}_2$, ${\bf  T}={\bf  s}_1-{\bf
s}_2$,
\begin{eqnarray}
\label{a_13} \hat{g}_{e}^{pn}(0)=G_{e}^{p}(0){I}_{1}({\bf
h})-G_{e}^{n}(0){I}_{2}({\bf h})\, ,\nonumber\\
\hat{g}_{m}^{pn}(0)=G_{m}^{p}(0){I}_{1}({\bf
h})-G_{m}^{n}(0){I}_{2}({\bf h})\, ,\nonumber\\
\hat{G}_{m}^{pn}(0)=G_{m}^{p}(0){I}_{1}({\bf
h})+G_{m}^{n}(0){I}_{2}({\bf h})\, ,\nonumber\\
\hat{G}_{e}^{pn}(0)=G_{e}^{p}(0){I}_{1}({\bf
h})+G_{e}^{n}(0){I}_{2}({\bf h})\, ,
\end{eqnarray}
\begin{equation}
%
w\equiv w({\bf q}) =\sqrt{m^2+{\bf q}^2},\nonumber
\end{equation}
\begin{eqnarray}
{I}_{i}({\bf h})\chi({\bf q})=
\chi({\bf d}_i({\bf  q})), \nonumber\\
{\bf d}_i({\bf q})={\bf q}+(-1)^i{\displaystyle \frac{2{\bf
h}}{1-h^2}}\,( w+(-1)^i({\bf h}\cdot {\bf q}) )\approx{\bf q}+(-1)^i
2{\bf h} w\,,
\end{eqnarray}
where $G_{m}^{n}(Q_N^2)$, $G_{m}^{p}(Q_N^2)$, $G_{e}^{n}(Q_N^2)$,
$G_{e}^{p}(Q_N^2)$ are nucleon electromagnetic
form factors parameterized according to Ref.~\cite{Kelly}. 

In Ref.~\cite{Allen} the elastic electron-deuteron scattering was
described in frames of the PF RKM. It was shown that in the PF SA
 that the
momentum of the unstruck particle (the spectator) is unchanged,
while the impulse given to the struck particle is not the impulse
given to the deuteron.

Following a general approach to construction of the electromagnetic
current operator for relativistic composite system \cite{Lev} we
define the momentum transfer $Q^2_i$ to the particle $i$ as an
increment of the particle 4-momentum $q_i$ \cite{Allen}%
\begin{equation}
Q^2_i=|(q_i'-q_i)^2|.
\end{equation}
For interacting particles the individual 4-momenta are not defined
before photon absorption  as well as after it.
Therefore we introduce an operator ${ Q}^2_i$ corresponding to the
physical quantity of  the momentum transfer ${Q}^2_i$. In Appendix B
we  generalize the deduction presented in Ref.~\cite{Allen} and show
that
\begin{equation}
Q^2_1=-(q_1'-q_1)^2=16(m^2+{\bf q}^2-\frac{({\bf q}\cdot{\bf h})^2}{h^2})\frac{h^2}{(1-h^2)^2}. \label{fatransfer}%
 \end{equation}
This is the general expression of the  ${ Q}^2_1={ Q}^2_2={ Q}^2_N$
in case of free two-particle states (for particles of equal masses),
therefore we use this expression in the PF SA for evaluation of the
current operator in Eq.~(\ref{preciseCurr}). The parameter ${\bf h}$
does not depend on interaction and is specified by relative
''position''\hspace{1mm} of initial $NN$ c.m. frame and final $NN$
c.m. frame. In case
 of two interacting particles ${\bf q}$ and $Q^2_i$ are
operators in the internal space. In impulse representation ${\bf q}$
is a variable of integration \cite{Allen}. It is obvious that in
action on a plane wave (for PlWA) this operator is equivalent to the
multiplication by a number ${ Q}^2_N>0$ if $h\neq 0$. The first
order in $h$ approximation gives $Q^2_N\approx0$ for
Eq.~(\ref{FSIcurrent}). Consideration similar to that of
Ref.~\cite{Allen} gives for the PlWA  in our case of the deuteron
photodisintegration
\begin{equation}
Q_N^2=E_{\gamma}^2-(w_n'-w_p')^2=(2w_n'-m_D)(2w_p'-m_D),\label{finaltransferMy}%
 \end{equation}
where $w_n'$ and $w_n'$ are final energies of neutron and proton in
the initial c.m. frame (laboratory frame). Detailed deduction of
Eq.~(\ref{finaltransferMy}) is published in Ref.~\cite{myTransfer}.

\section{\label{sec:IV}RESULTS AND FUTURE PROSPECTS}
\begin{figure*}[htbp!]
\includegraphics[width=8cm,height=7.5cm]{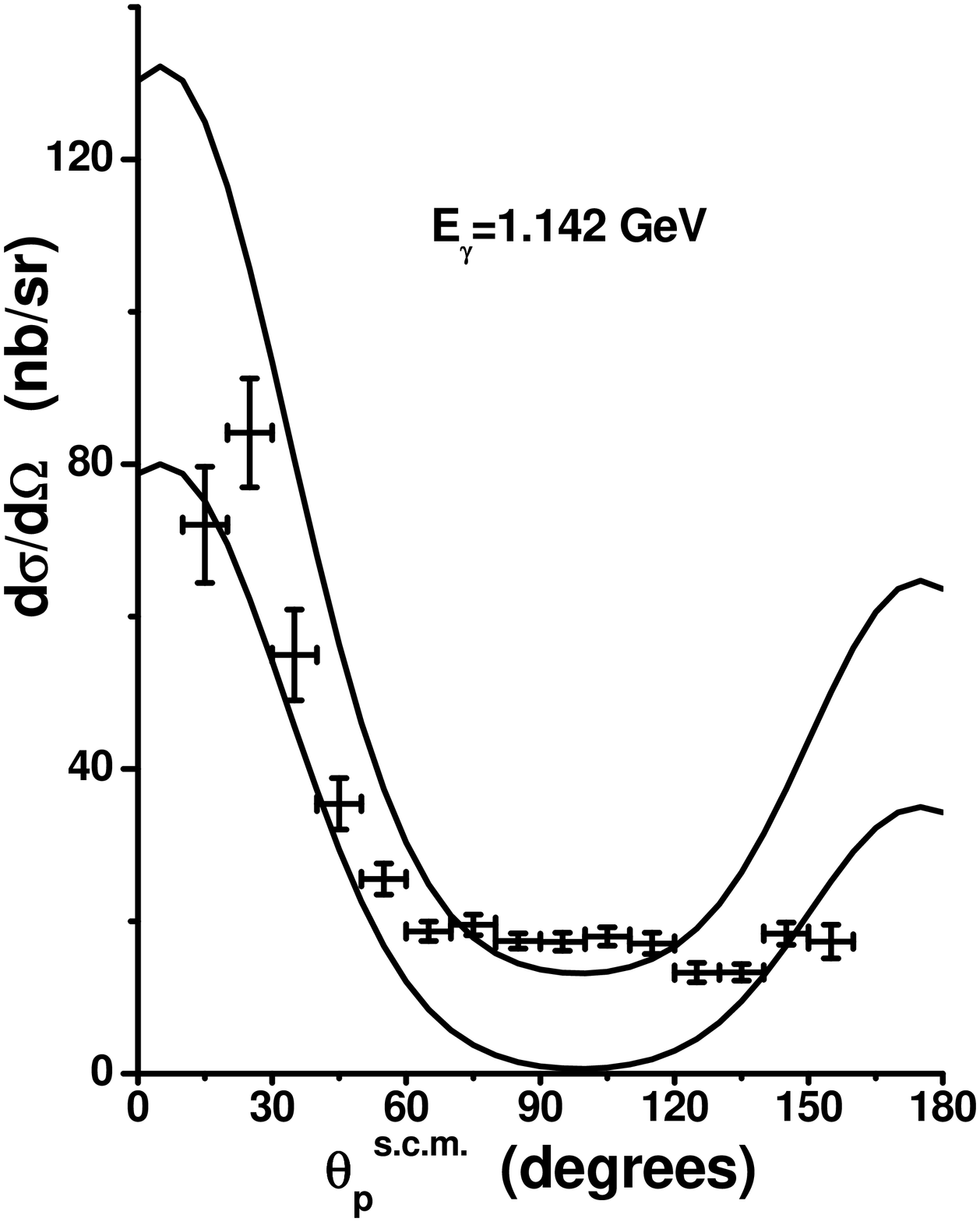}    \includegraphics[width=8cm,height=7.5cm]{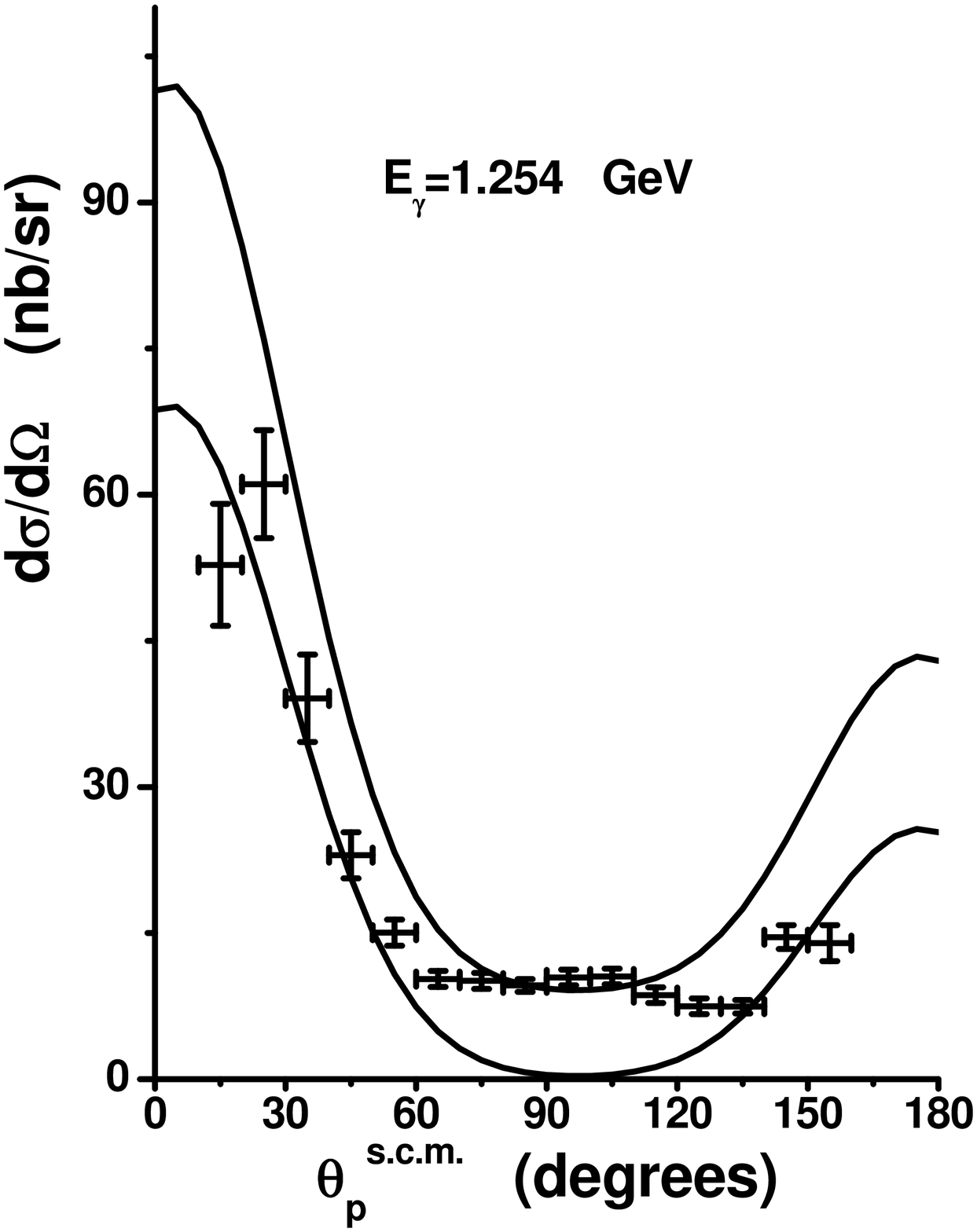}
\includegraphics[width=8cm,height=7.5cm]{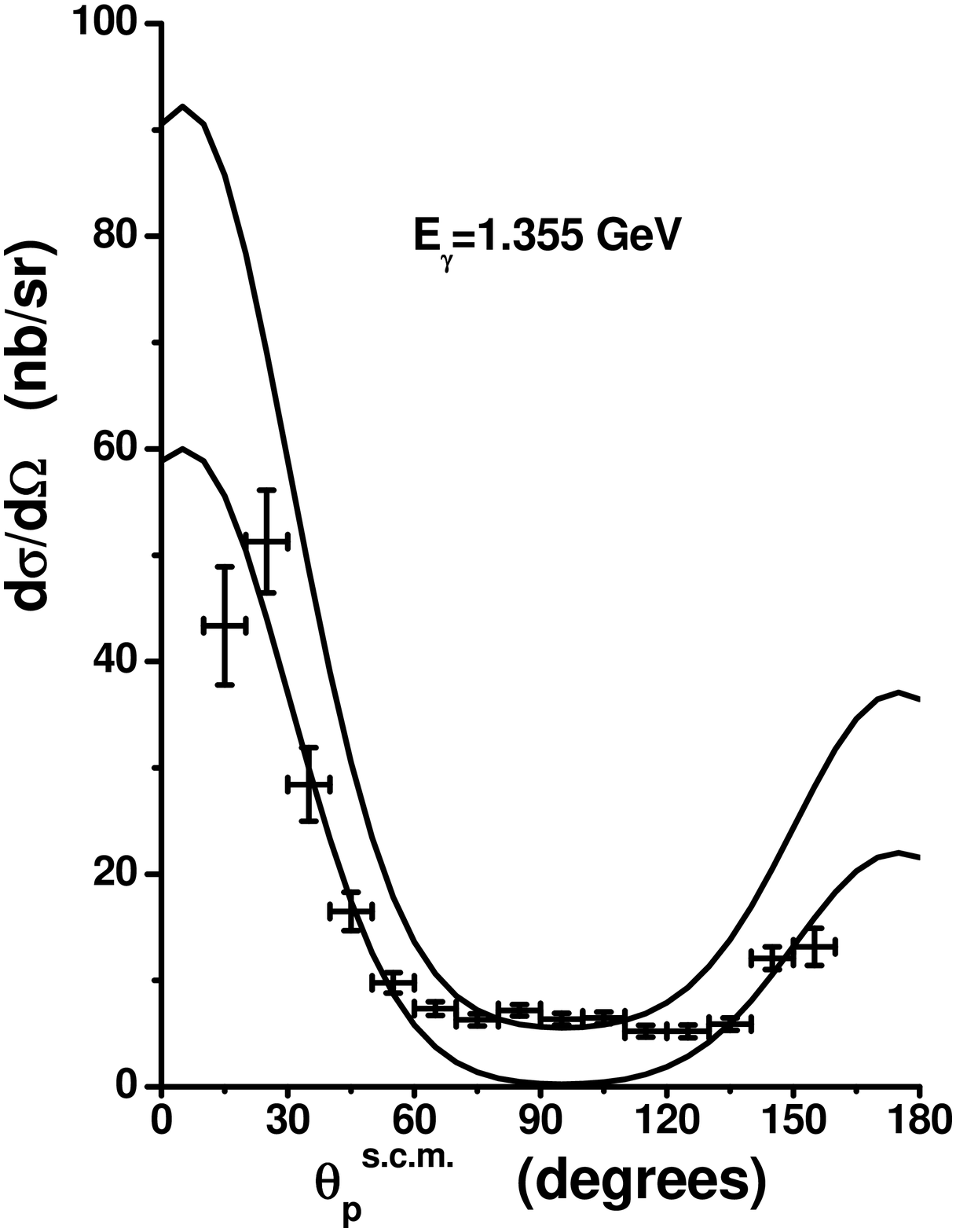}    \includegraphics[width=8cm,height=7.5cm]{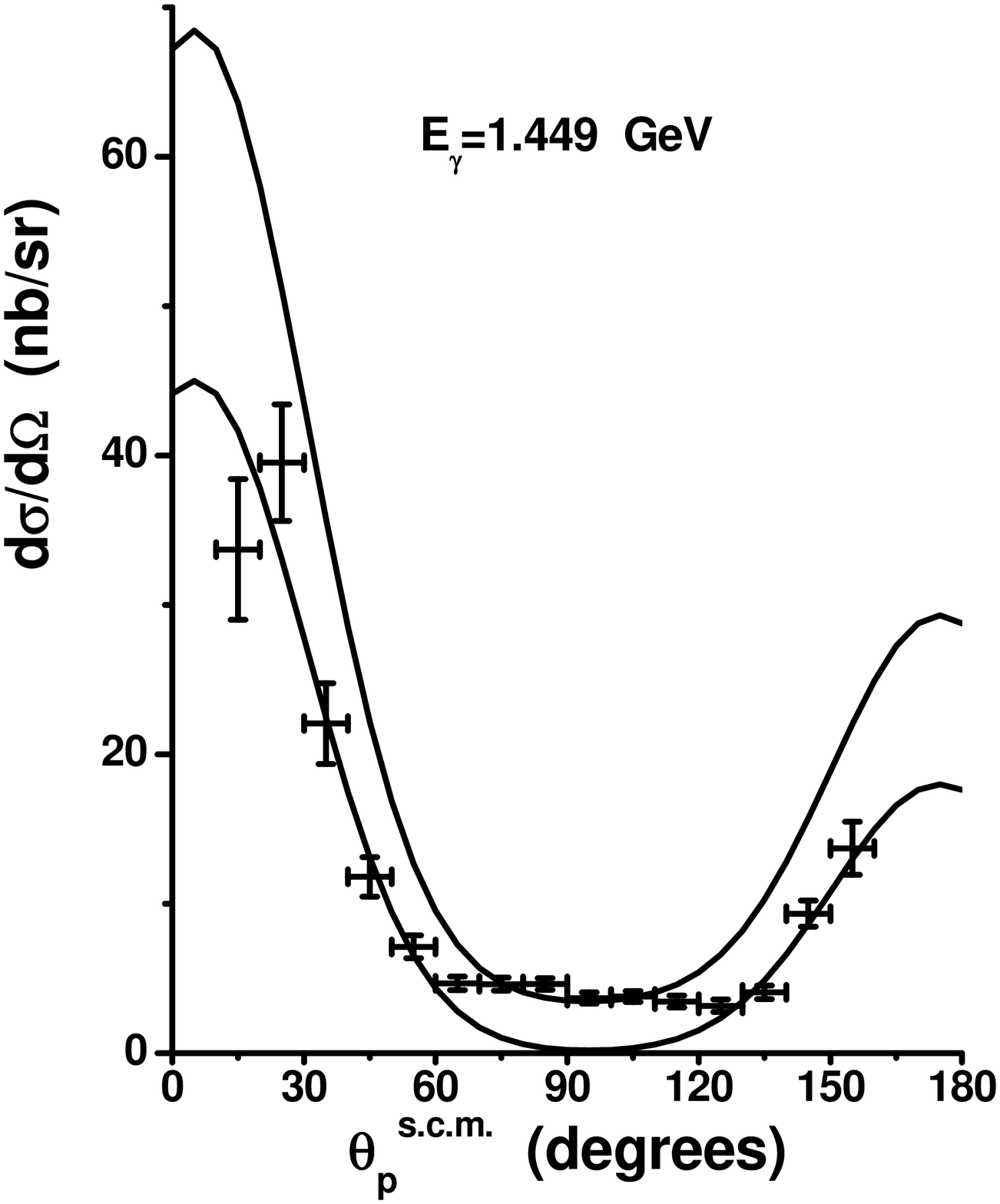}
\includegraphics[width=8cm,height=7.5cm]{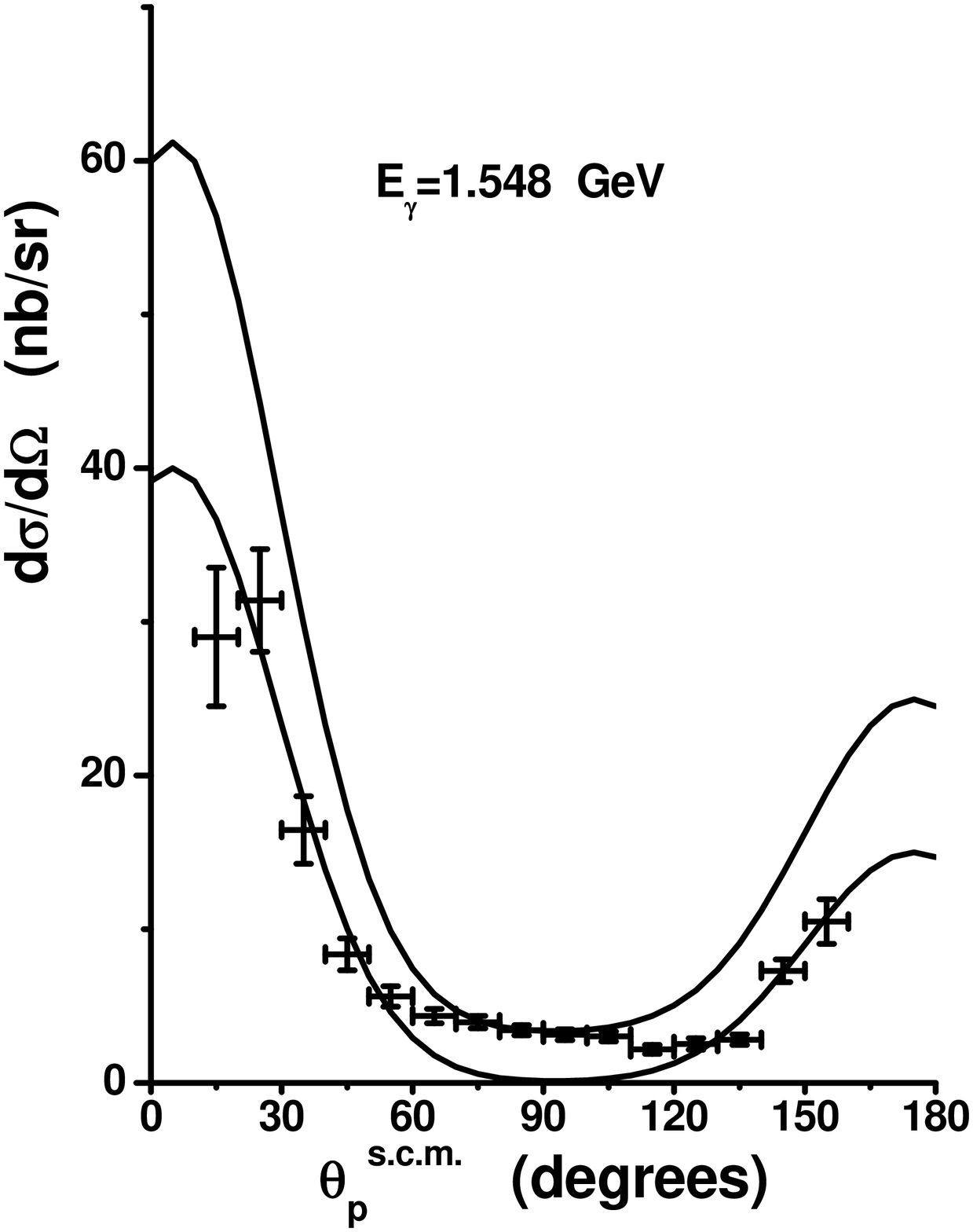}    \includegraphics[width=8cm,height=7.5cm]{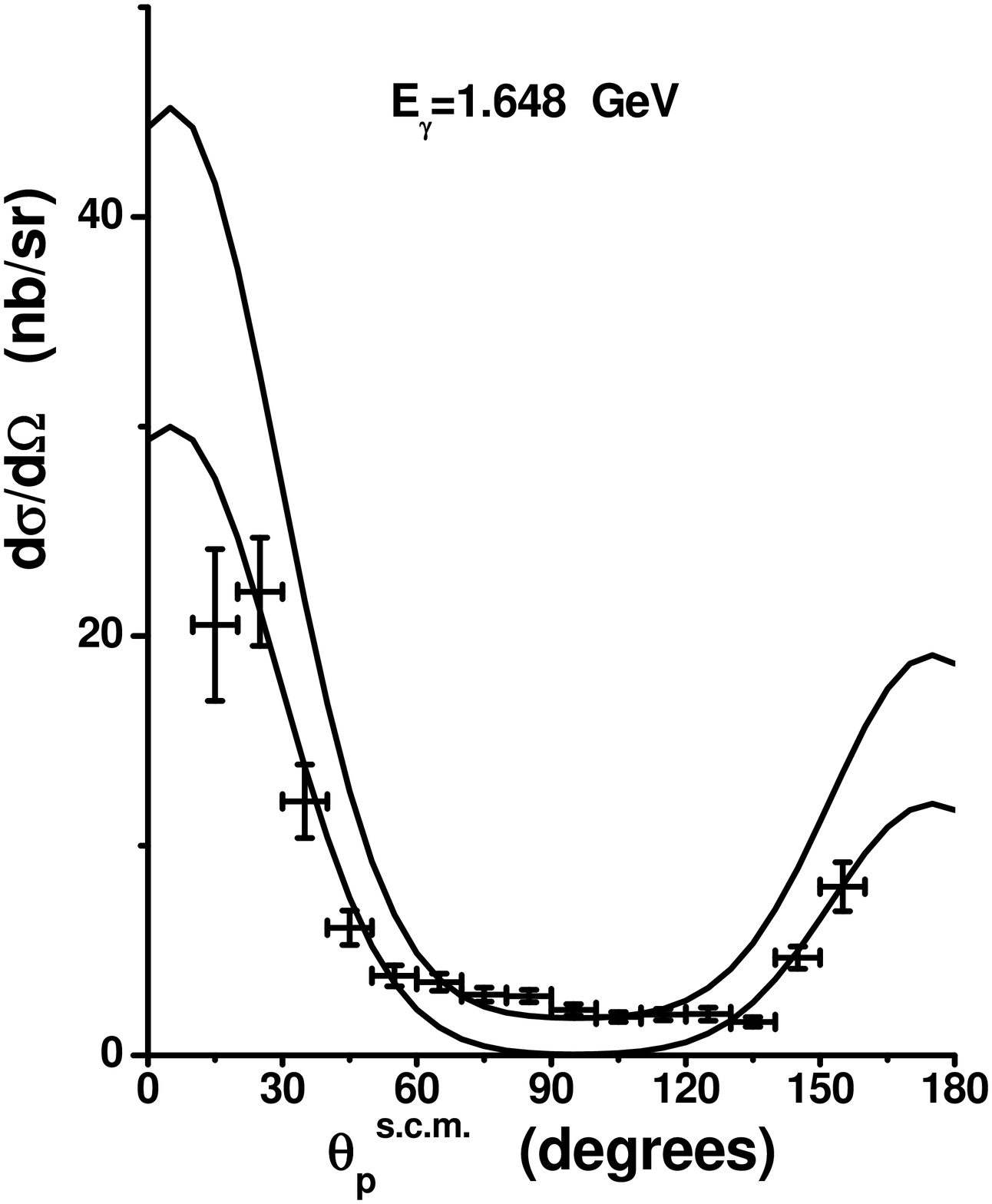}
\caption{\label{fig6} Angular dependence of $d\gamma\to pn$ reaction
differential cross-sections for different photon energies
$E_{\gamma}$. Our theory is compared with experiment \cite{N4}.}
\end{figure*}

\begin{figure*}[htbp!]
\includegraphics[width=8cm,height=7.5cm]{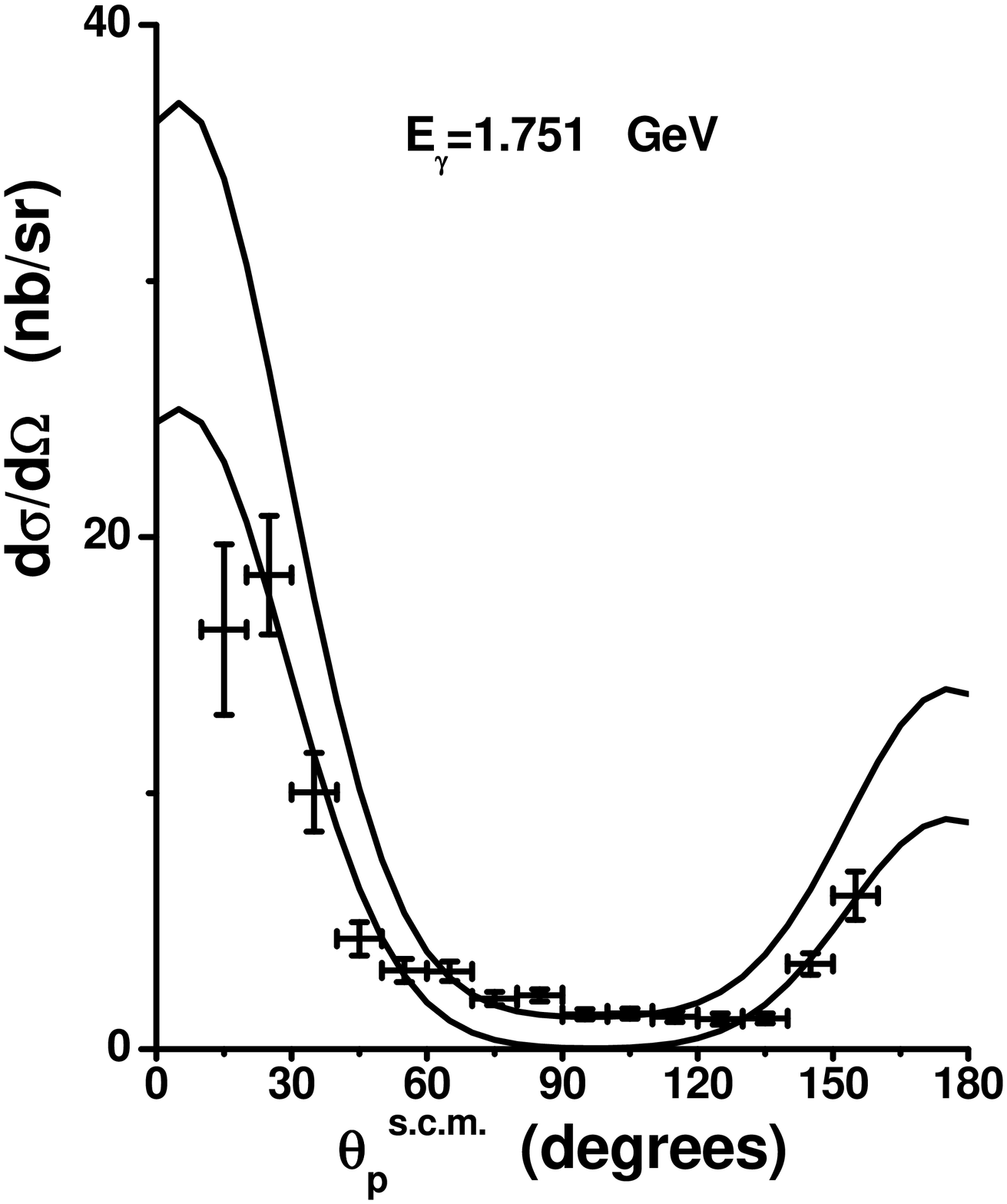}    \includegraphics[width=8cm,height=7.5cm]{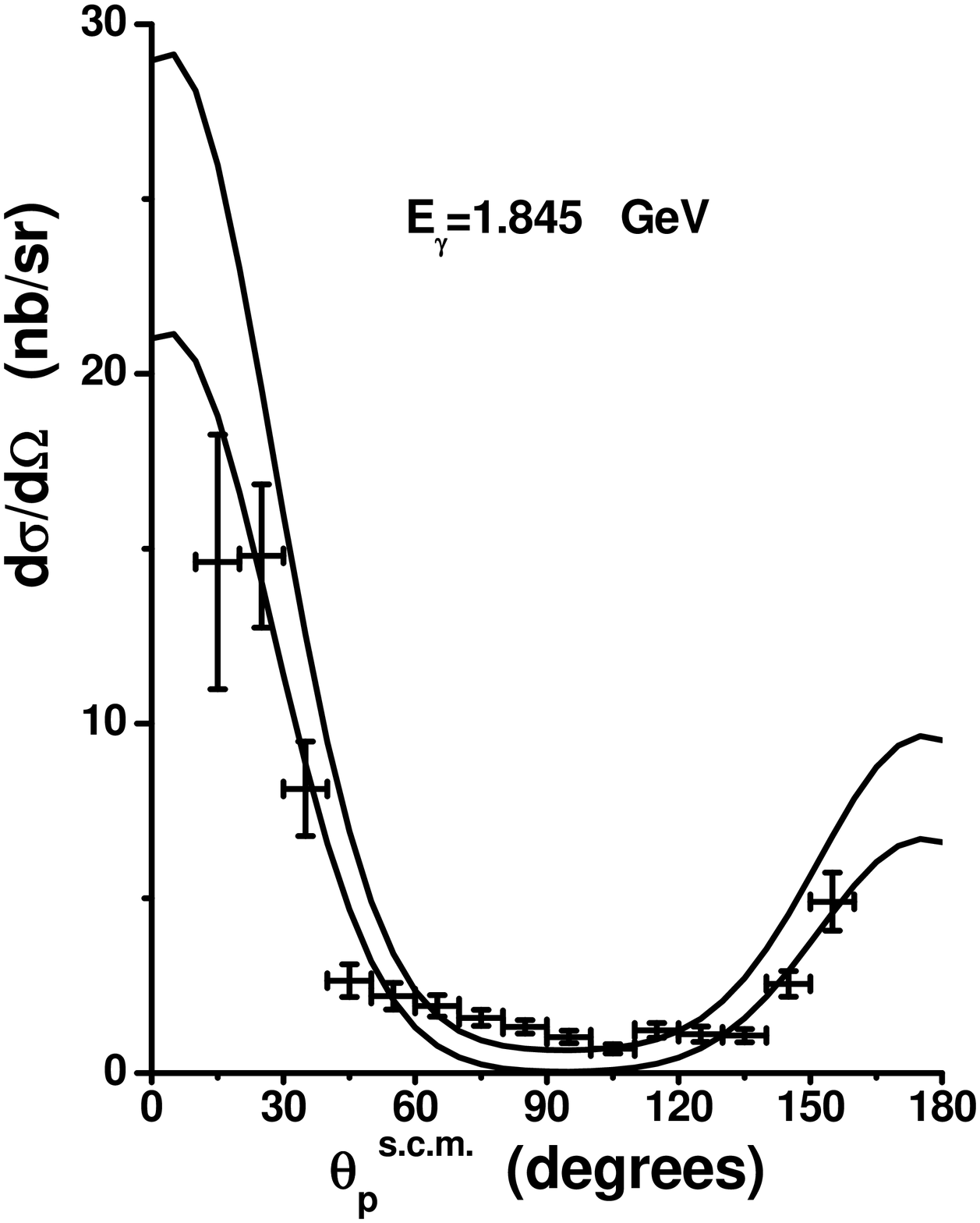}
\includegraphics[width=8cm,height=7.5cm]{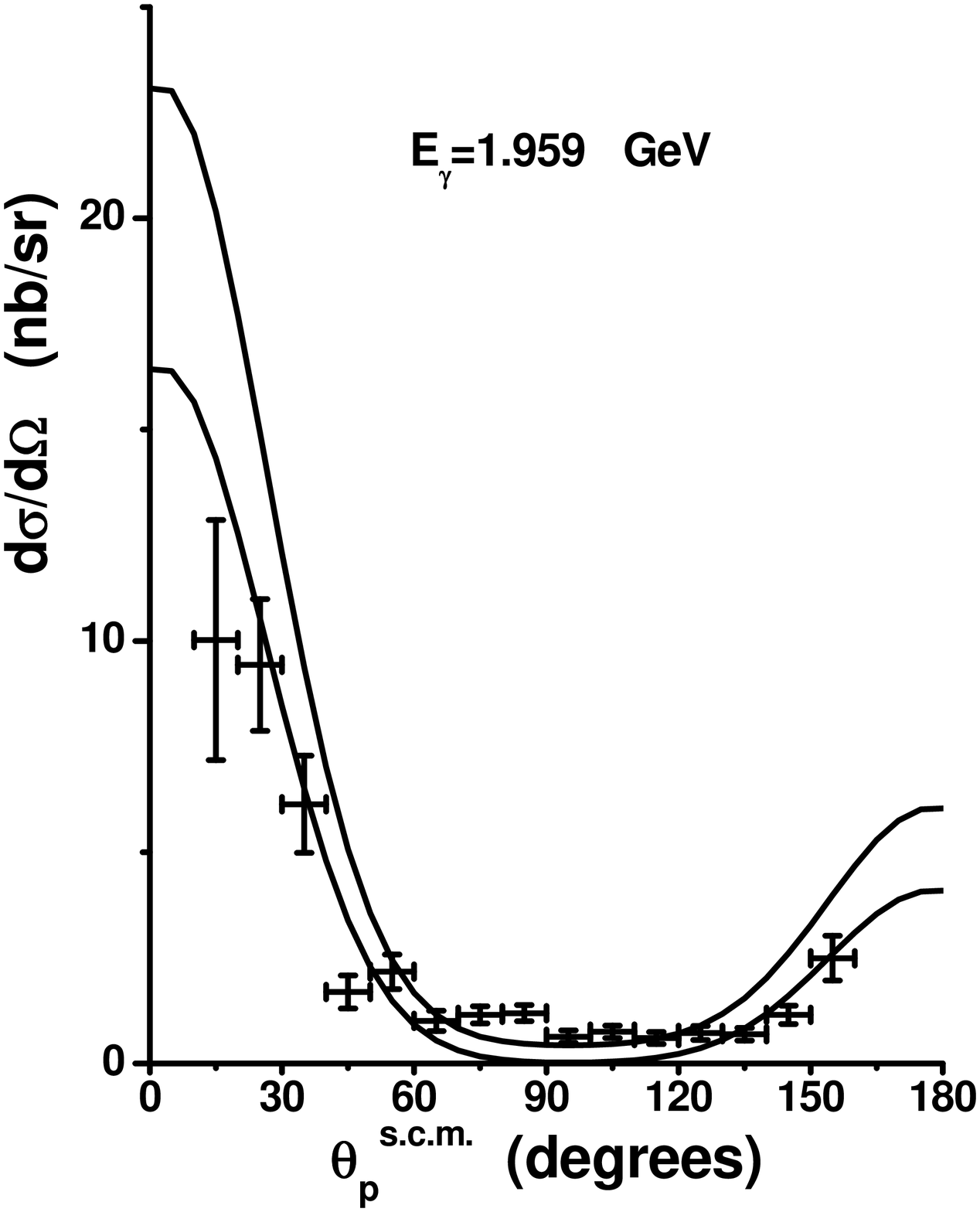}     \includegraphics[width=8cm,height=7.5cm]{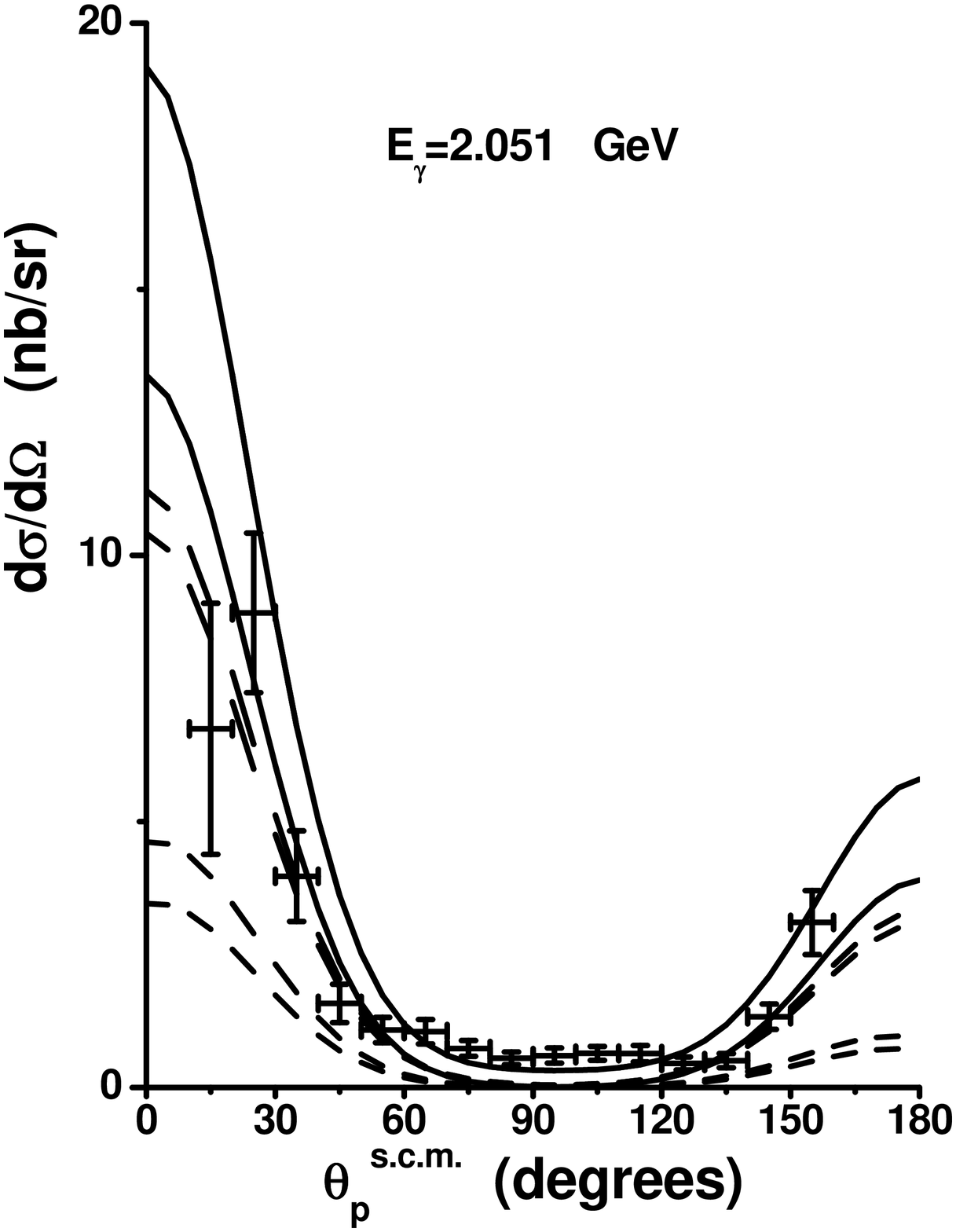}
\includegraphics[width=8cm,height=7.5cm]{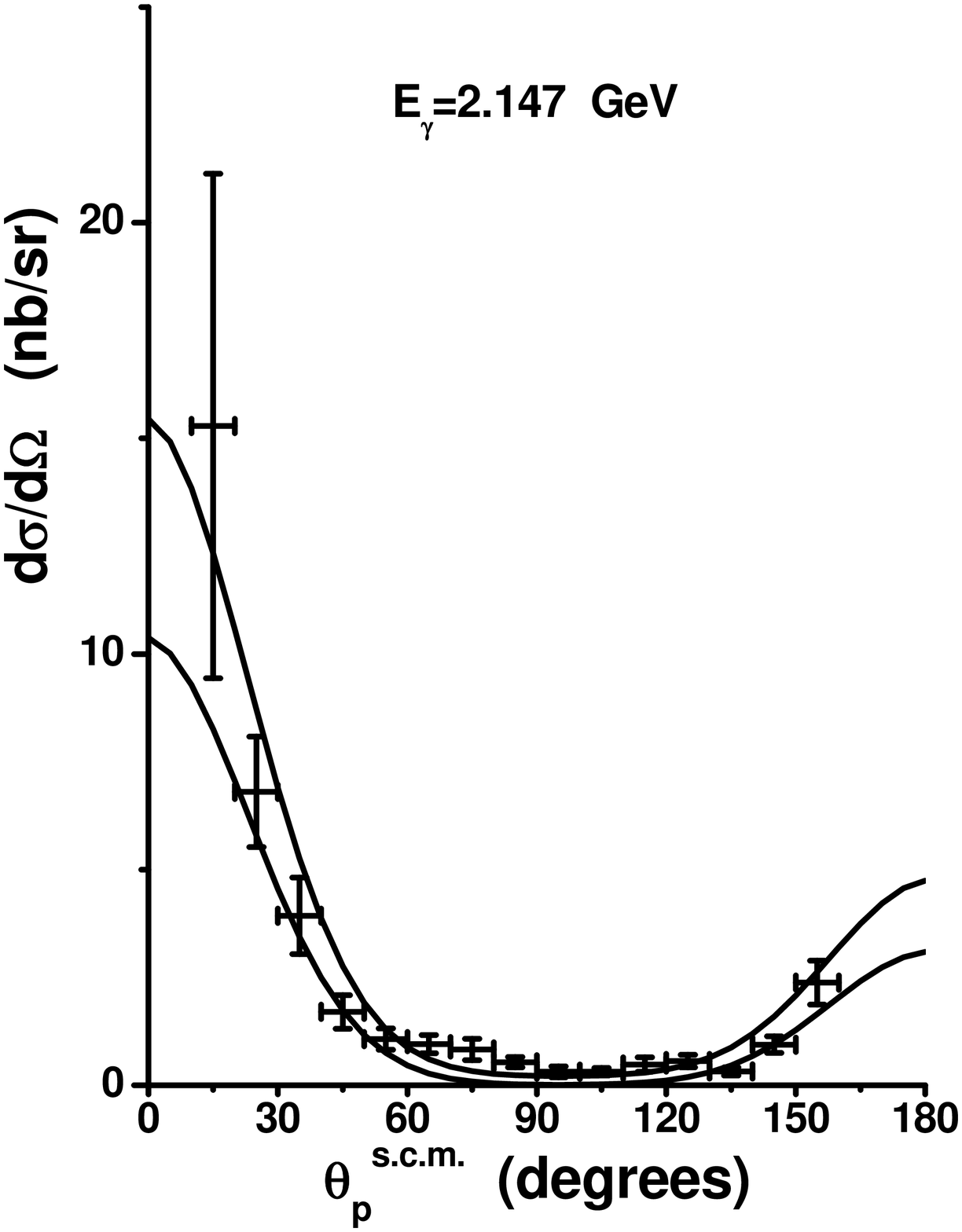}    \includegraphics[width=8cm,height=7.5cm]{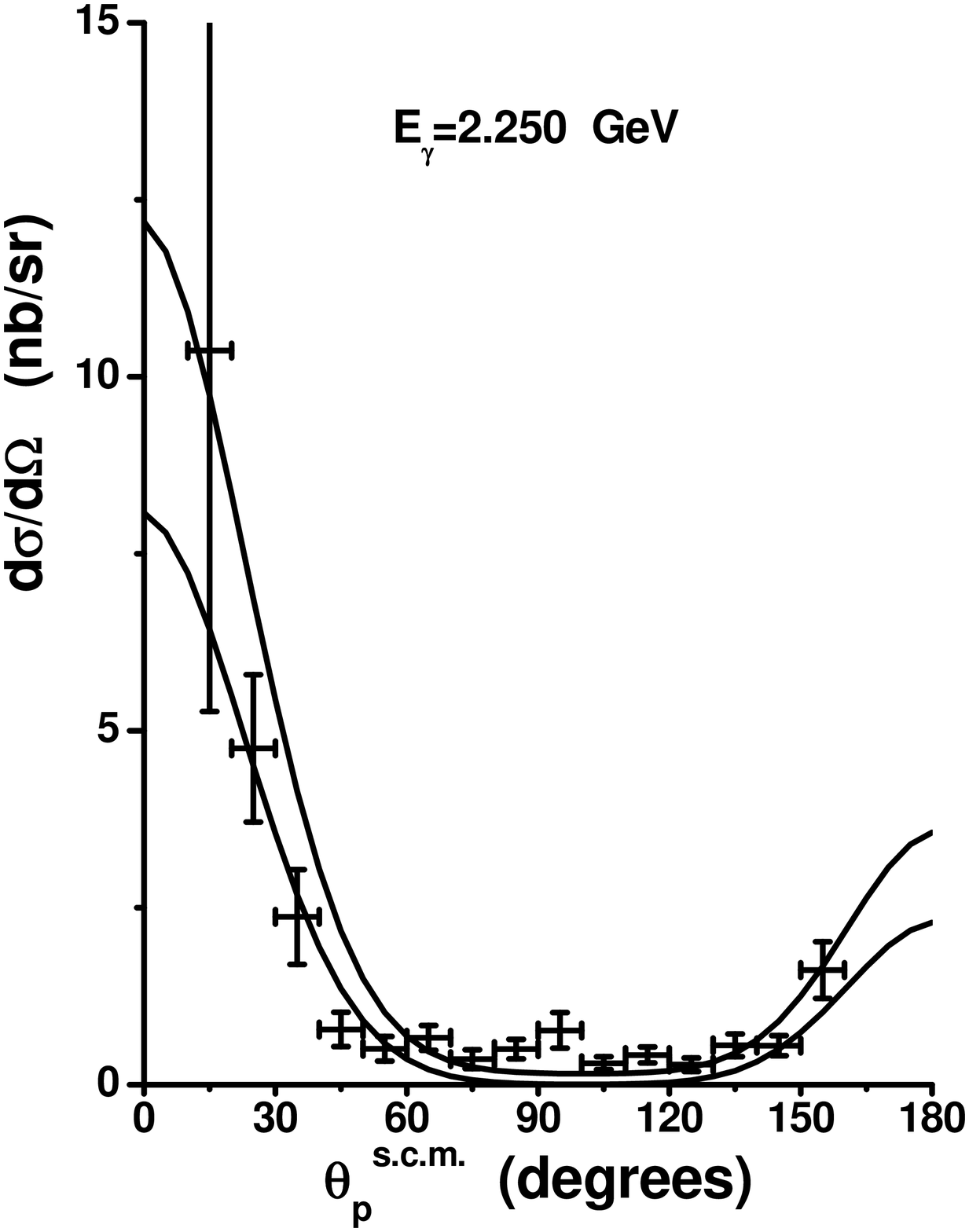}
 \caption{\label{fig:crossParis}  The same as in Fig. \ref{fig6}, but (for $E_{\gamma}=2.051$~GeV)
 our theoretical results for the RCP (Paris
 potential \cite{Paris})
is also shown (two lower dashed curves). Upper dashed curves show
results of our calculations for PlWA  with deuteron wave function in
the initial state calculated with our MP. }
\end{figure*}

\begin{figure*}[htbp!]
\includegraphics[width=8cm,height=7.5cm]{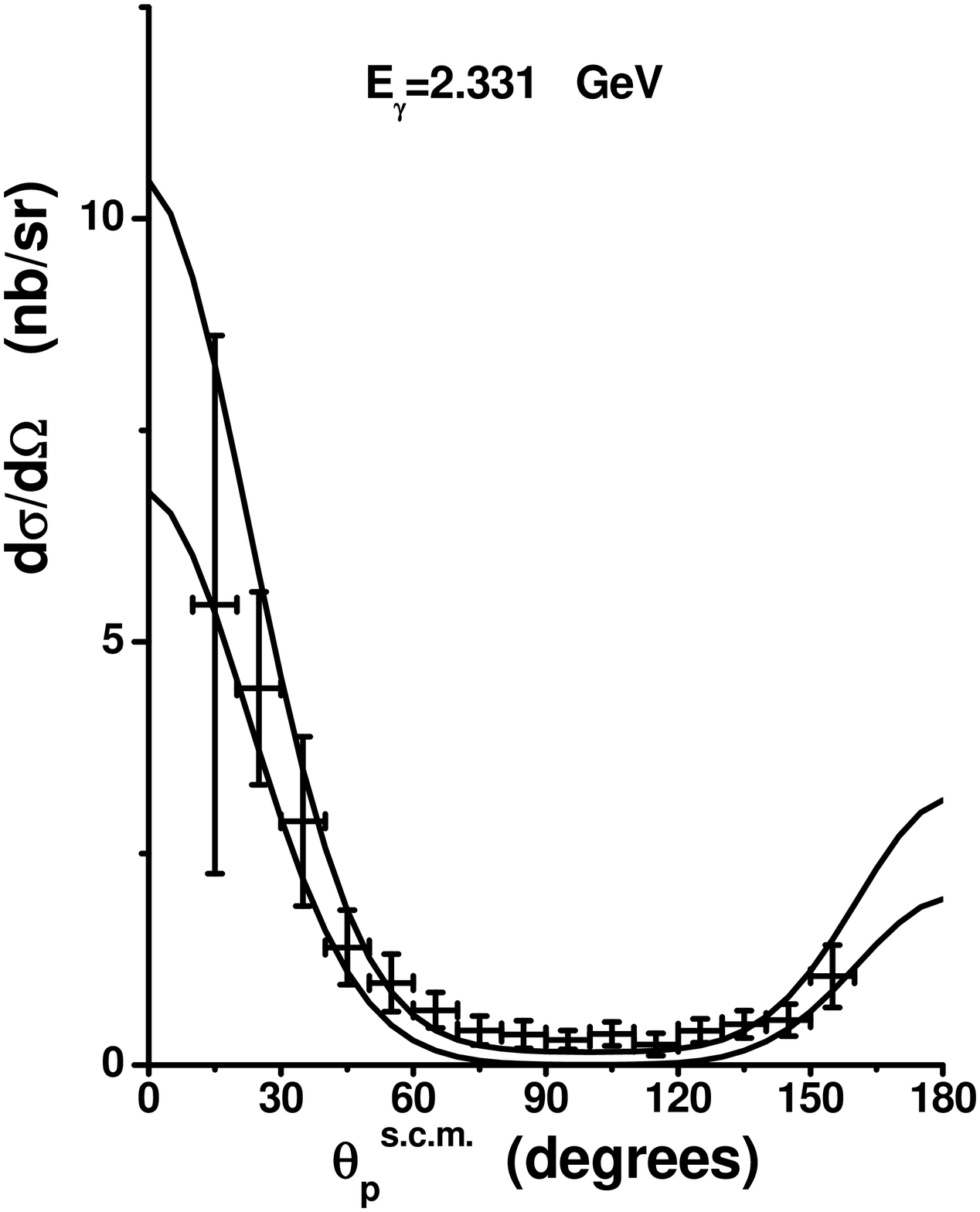} \includegraphics[width=8cm,height=7.5cm]{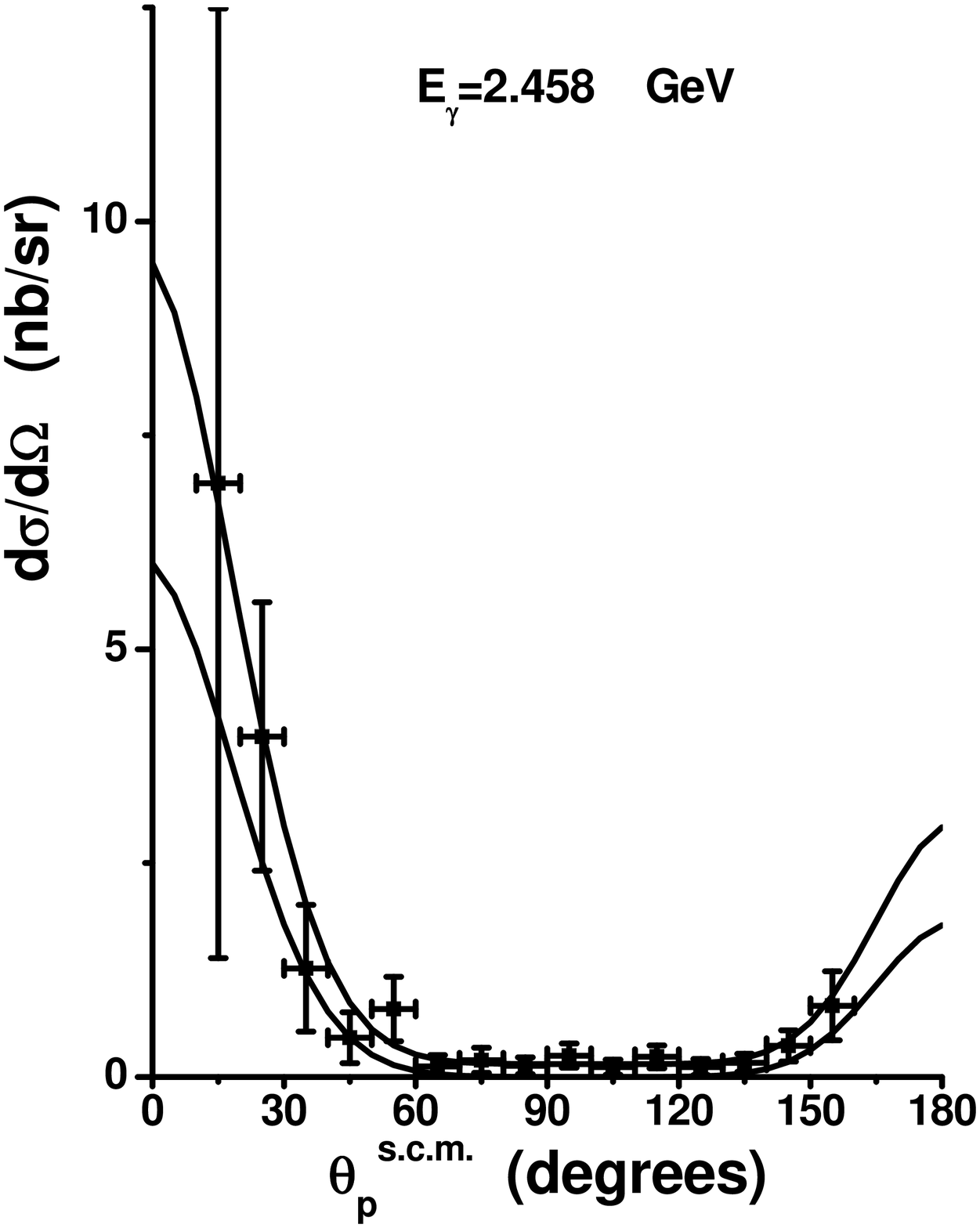}
 \caption{\label{fig8} The same as in Fig. \ref{fig6} }
\end{figure*}

Our theoretical description of the differential cross-section of
$d\gamma\to pn$ reaction is compared with recent experiment
\cite{N4} in Figs.~\ref{fig6}-\ref{fig8}  at a few energies around
$E_\gamma =2$ GeV. We do not use free parameters. However, there are
uncertainties in our calculation. First uncertainty  is caused by
uncertainty in the form factor parametrization of Ref.~\cite{Kelly}
due to  errors of the experimental data on the form factors. We
estimated this uncertainty at about $\pm 15$ per cent of the results
varying parameters inside limits defined in Ref.~\cite{Kelly}.
Second uncertainty is connected with the approximation
(\ref{FSIcurrent}) used to calculate the FSI term in
Eq.~(\ref{matixElAll}). In Eq.~(\ref{FSIcurrent}) the nucleon form
factors are equal to their values at
 $Q^2_N=0$ and therefore the FSI term  is overestimated.
  Fig.~\ref{fig:crossParis} ($E_{\gamma}=2.051$~GeV) shows contribution
 of the FSI term.  The PlWA term of the
amplitude (\ref{matixElAll}) is dominant, but the FSI is not
negligible. Therefore
 it is desirable
 to estimate  the second order in $h$
correction to the approximation (\ref{FSIcurrent}). We plan to do
this estimation in the future. Third uncertainty is caused by
uncertainty of the addend $\delta {\bf j}$ that restores the current
conservation equation. To estimate this uncertainty we calculated
two curves for every energy of the photon.
 The lower curves
correspond to calculations  without addend $\delta {\bf j}$ in the
current operator and with form factors of Ref.~\cite{Kelly} varied
to their lower limits. The upper curves correspond to full
calculations
 and
with form factors varied to their upper limits. The FSI is included
for both curves.


\begin{figure}[htbp!]
\includegraphics[width=8.6cm,height=7cm]{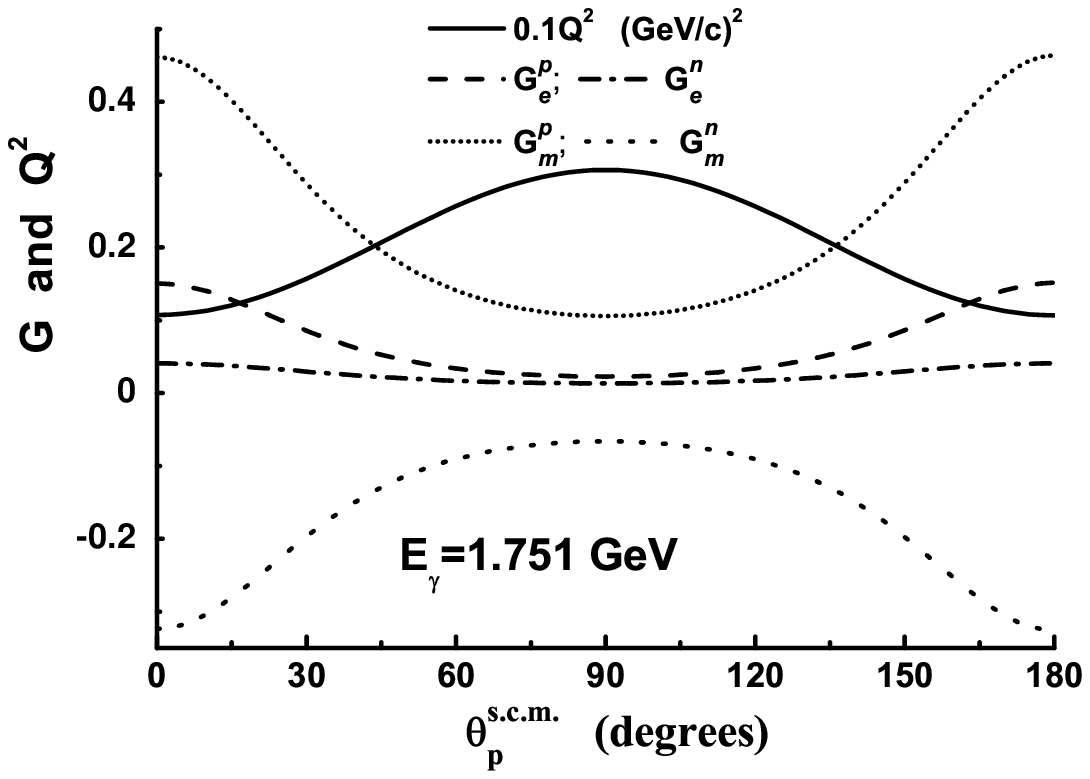}
 \caption{\label{fig:formfactors} Angular dependence of 4-momentum transfer $Q^2$, and of the nucleon electromagnetic
 form factors
 for $d\gamma\to pn$ reaction calculated from Eq.~(\ref{finaltransferMy}) for the PlWA.
 In our calculations we use dependance of the
 form factors on $Q^2$  according to parametrization of J.J.
 Kelly
 \cite{Kelly}.}
\end{figure}

We see good general correspondence of the theory and experiment both
in absolute values and in shape of angular dependence of the
differential cross-section at various energies. Large absolute
values of cross-sections in our theory in comparison with results
for the RCP
(Fig.~\ref{fig:crossParis} ($E_{\gamma}=2.051$~GeV)) originate
mainly in the nodal character of the deuteron $S$-wave functions
(greater weight of the high-momentum wave function components).
 The ability to describe both the
absolute value and the angular dependence of differential
cross-sections confirms the detailed algebraic structure of our
theory.  A persistent forward-backward asymmetry is determined
mainly by the angular dependence of the nucleon electromagnetic form
factors according to Fig.~\ref{fig:formfactors} (proton knockout
dominates at forward angles and neutron knockout dominates at
forward angles).

 To complete this line of our
investigation we plan to make an analysis of polarization
$d\gamma\to pn$ experiments and to consider the pionic radiative
capture $pp\to d\pi^{+}$ at proper energies. Other actual problems
are outlined in Ref. \cite{Myppg2}. First of them concerns the
microscopic theory of the MP. As we suppose, it is connected to the
short-range quark exchange between nucleons accompanied by
excitations of color dipole states of two virtual baryons with very
strong attraction between them. This scenario is based on the quark
configuration $s^4 p^2 \lbrack 42\rbrack_x$ in deuteron.

As the concluding remark it should be stressed that usage of the MP
instead of an RCP  in the theory of complex nuclei demands accurate
evaluation of $3N-$forces. Effect of these forces is much enhanced
\cite{N27}, as far as three nucleons without $NN$ core can overlap
and form short-range $9q$-subsystems with large probability. Recent
experiments \cite{N28} on the knock-out of nucleon from $^3 He$
nucleus may clarify the situation. In these experiments the missing
momentum is great and recoil to $2N$-subsystem with large relative
momentum of two spectator-nucleons is observed.

Authors are grateful to Prof. V.I.~Kukulin and to Dr.~N.A.~Burkova
for useful remarks.
\appendix
\section{\label{sec:APPENDIX}}

In this Sec. we explain calculation of the electromagnetic current
matrix elements. The derivation is based on results of
Ref.~\cite{Lev}, where Eq.~(\ref{MA_of_CO}) and
Eq.~(\ref{exact_current}) were deduced.

Let us define a matrix
~\cite{Moussa}
\begin{equation}
\alpha (g)= \frac{g^0+1+ {\bf  \sigma }\cdot {\bf g}}{
\sqrt{2(g^0+1)}},\label{alpha_g}
\end{equation}
corresponding to a 4-velocity $g$, where ${\bf \sigma} =( \sigma_x,
\sigma_y, \sigma_z)$ are the Pauli matrices. Let us define  the
matrix $\breve{p}=\widetilde{M}(p)\equiv \sigma^\mu p_\mu$
corresponding to a 4-vector $p$ ($\sigma^{0}$ is $2\times 2$ unit
matrix). Operator $\widetilde{M}(p)$ transforms the 4-vector $p$ to
$(2\times 2)$ matrix. The inverse transformation is defined as
\begin{eqnarray}
&p_0=& \frac{1}{2}( \breve{p}_{11}+ \breve{p}_{22}),\ \  p_1=
\frac{1}{2}( \breve{p}_{12}+ \breve{p}_{21}),\nonumber\\%
&p_2=&\frac{1}{2i}(\breve{p}_{21}- \breve{p}_{12}),\  \ p_3=
\frac{1}{2}( \breve{p}_{11}-
\breve{p}_{22})\label{vector_from_matrix},
\end{eqnarray}
we denote this transformation as $p=\widetilde{V}(\breve{p})$. The
boost  $p \to L[\alpha (g)]p$ is equivalent to matrix transformation
\begin{equation}
\breve{p} \to \alpha (g) \breve{p} \alpha (g)^+.
\end{equation}
It is easy to see that $L[\alpha (g)](1,0,0,0)=g$. The Poincar\'{e}
group transformation $U(a,l)$ is characterized ~\cite{Lev} by the
4-shift $a$ and 4-rotation $l$:
\begin{equation}
U( \alpha ,l) \varphi (g) = e^{\imath mg'a} D[ {s}; \alpha (g)^{-1}l
\alpha (g')] \varphi (g'),
\end{equation}
where $\varphi (g)$ is a normalized spinor function of a particle
with mass $p$; ${s}$ is spin of the particle; and $g'=L(l)^{-1}g$.
In our case of spin $s$ = 1/2 particles, we deal with the
fundamental representation ~\cite{Moussa}, i.e. ${\bf s}_i\equiv
\frac{1}{2}{\bf \sigma}_i$ and
\begin{equation}
D( {s}; \alpha (g)^{-1}l \alpha (g'))\equiv \alpha (g)^{-1}l \alpha
(g').
\end{equation}

The ''internal'' electromagnetic current operator for a system of
two particles in the SA is~\cite{Lev}
\begin{equation}\label{exact_current}
 j^\mu( {\bf h})= \sum_{i=1,2} (L^{i})_\nu ^\mu
D^{i}_{1}D^{i}_{2}j_i^\nu ( {\bf h})D^{i}_{3}K^{i}I_i( {\bf h}),
\end{equation}
where
\begin{equation}
(L^{i})_\nu ^\mu= L \left( L[ \alpha (f)] \frac{q_i}{m_i}, L[ \alpha
(f')] \frac{d_i}{m_i} \right)_\nu ^\mu,
\end{equation}
\begin{eqnarray}
D^{i}_{1}=D \left[ {s}_k; \alpha ({q_k}/{m_k})^{-1} \alpha (f)^{-1}
 \alpha (f') \alpha (d_{ki}/m_k) \right]=\nonumber\ \ \ \ \ \ \\
 =\alpha_k (q_k/m_k)^{-1} \alpha_k (f)^{-1}
 \alpha_k (f') \alpha_k (d_{ki}/m_k),\ \ \ \ \ \ \
 \end{eqnarray}
\begin{eqnarray}
D^{i}_{2}=D \left[ {s}_i; \alpha (q_i/m_i)^{-1} \alpha (f)^{-1}
\alpha \left(z_{i} \right) \right]=\nonumber\ \ \ \ \ \ \\ =\alpha_i
(q_i/m_i)^{-1} \alpha_i (f)^{-1} \alpha_i \left(z_{i} \right),\ \ \
\ \ \ \
\end{eqnarray}
\begin{eqnarray}
D^{i}_{3}=D \left[ {s}_i; \alpha (f'_i)^{-1} \alpha \left(z_{i}
\right)^{-1} \alpha (f') \alpha (d_i/m_i) \right]=\nonumber\ \ \ \ \
\ \\ =\alpha_i (f'_i)^{-1} \alpha_i \left(z_{i} \right)^{-1}
\alpha_i (f') \alpha_i (d_i/m_i),\ \ \ \ \ \ \
\end{eqnarray}
kinematical multipliers
\begin{equation}
K^{i}=\frac{m_iw_i( {\bf q}_i)}{w_i( {\bf d}_i)} \left( \frac{M(
{\bf d}_i)}{M( {\bf q})} \right)^{3/2}\label{exact_current_ended}
\end{equation}
 Here, $k$ =
2, if $i$ =1, and, conversely, $k$ = 1, if $i$ = 2.
 $L(G,G')$ denotes the Lorentz transformation $L[\alpha(G,G')]$, and $\alpha(G,G')\equiv\alpha((G+G')/|G+G'|)$,
 $z_{i}= L[
\alpha (f)] q_i/m_i,L[ \alpha (f')] d_i/m_i$. Next,
\begin{equation}
f=L(G,G')^{-1}G, \qquad f'=L(G,G')^{-1}G'
\end{equation}
represent the 4-velocities of the two-nucleon c.m. in the initial
and final state, respectively, meaning the coordinate frame
(\ref{LevFrame}). The following formal aspects should be mentioned
here: $f^2=f'^2=1$, ${\bf f}+ {\bf f}'=0$, $f^0=f^{0'}= (1+ {\bf
f}^2)^{1/2}$, ${\bf h}\equiv {\bf f}/f^0$; $L(G,G')=L( \alpha
(G,G'))$, $\alpha (G,G') =\alpha ((G+G')/|G+G'|)$; $d_1=(w_1( {\bf
d}_1), {\bf d}_1)$, $d_2=(w_2( {\bf d}_2), {\bf d}_2)$, $d_{12}=L[
\alpha (f')^{-1} \alpha (f)]q_2=(w_2( {\bf d}_1),- {\bf d}_1)$,
$d_{21}=L[ \alpha (f')^{-1} \alpha (f)]q_1=(w_1( {\bf d}_2), {\bf
d}_2)$; last equations gives $d_i$ also. Index $i$ or $k$ of
matrices $\alpha$ and $\sigma$ means that it acts in $i$-th or
$k$-th particle spin space and appears as in Eq.~(\ref{alpha_g}) but
with $\sigma_i$ or $\sigma_k$ correspondingly instead of $\sigma$.
Let $d_1=(\omega_1({\bf d}_1),{\bf d}_1)$, $d_2=(\omega_2({\bf
d}_2),-{\bf d}_2)$ and $I_i({\bf h})$ $(i=1,2)$ be operators defined
by the conditions $I_i({\bf h})\chi({\bf q})=\chi({\bf d}_i)$.
$g_{i}'=L[\alpha(f)]\frac{q_{i}}{m_{i}}$,
$g_{i}''=L[\alpha(f')]\frac{d_{i}}{m_{k}}$,
$f_{i}=L[z_{i}]^{-1}g_{i}'$, $f_{i}'=L[z_{i}]^{-1}g_{i}''$, ${\bf
h}_{i}=\frac{{\bf f}_{i}}{f_{i}^{0}}$,
$w_i(q)\equiv\sqrt{m_i^2+q^2}$.

Finally, $j_i^\nu ({\bf h})$ is a 4-current of the particle $i$,
$$
j_i^0 ( {\bf h})=eF_e^i \left(Q_i^2 \right),
$$

\begin{equation}
{\bf j}_i ( {\bf h})=- \frac{ie}{ \sqrt{1- {\bf h}^2_i}}F_m^i
\left(Q_i^2 \right) ( {\bf h}_i \times {\bf
s}_i),\label{one_current}
\end{equation}
where vectors ${\bf h}_i$ are defined below, ${\bf s}_i\equiv {\bf
\sigma}_i/2$, $Q_i^2={4m_i^2{\bf h}^2_i}/{\sqrt{1- {\bf h}^2_i}}$
(see also Eq.~(\ref{fatransfer})).

From (\ref{exact_current}-\ref{one_current}) it is obvious that for
a plane wave final state, when operator  ${\bf  q}=-i{\bf \nabla}$
can be substituted by vector ${\bf q}_f$, operator ${j}({\bf h})$
becomes an exterior product $j^{\nu}({\bf h})\equiv
\sum_{i=1,2}A_{i\nu}^i \otimes A_{k\nu}^i\,I_i( {\bf h})$; $k$ = 2,
if $i$ =1, and, conversely, $k$ = 1, if $i$ = 2. The ${\bf
q}_f$-dependent matrix $A_{i\nu}^k$ acts in $i$-th particle spin
space and presentation $A_{i\nu}^k=\sigma^{\mu}_i
a^k_{i\nu\mu}\equiv a^k_{i\nu 0}+2({\bf s}_i\cdot {\bf a}^k_{i\nu})$
is valid. ''Components'' $a_{\nu}$ are extracted by
(\ref{vector_from_matrix}) transformation
\begin{eqnarray}
a_{\nu i}^i=\widetilde{V}((L^{i})_\mu ^\nu D^{i}_{2}j_i^\mu (
{\bf h})D^{i}_{3}K^{i})\nonumber\\
a_{\nu i}^k=\widetilde{V}(D^{i}_{1}),\ \ i\neq k.
\end{eqnarray}
Functions $B$ of Eq.~(\ref{preciseCurr}) are expressed as
\begin{eqnarray}
{B}^{\nu}_{1i}=a^{i}_{i\nu 0}a^{k}_{i\nu 0},\ {\bf
B}^{\nu}_{2i}=2a^k_{i\nu 0}{\bf a}^i_{i\nu},\
{\bf B}^{\nu}_{3i}=2a^i_{i\nu 0}{\bf a}^k_{i\nu},\ \nonumber\\
{\bf B}^{\nu}_{4i}=2{\bf a}^i_{i\nu},\ {\bf B}^{\nu}_{5i}=2{\bf
a}^k_{i\nu},\nonumber\ \ \ \ \
\end{eqnarray}
$k$ = 2, if $i$ =1, and, conversely, $k$ = 1, if $i$ = 2.

Now, we should take into account the current conservation equation
\begin{equation}
\frac{\partial\hat{J}^\mu (x)}{\partial x^\mu}=0.
\end{equation}
Using also the 4-shift
\begin{equation}
\hat{J}^\mu (x)=exp(i \hat{P}x) \hat{J}^\mu (0) exp (-i
\hat{P}x),\label{J0}
\end{equation}
we obtain a relation
\begin{equation}
 \hat{P}_\mu \hat{J}^\mu (0)-\hat{J}^\mu (0)\hat{P}_\mu =0.\label{CCE2}
\end{equation}

In terms of the internal variables of $NN$-system, Eq.~(\ref{CCE2})
can be reduced to the matrix element
\begin{equation}
< \chi_f| {M}_{f}G_{f\,0}j^{0}({\bf h})-{M}_f{\bf G}_{f}{\bf j}({\bf
h})- {M}_iG_{i\,0}j^{0}({\bf h})+{M}_i{\bf G}_{i}{\bf j}({\bf h})|
\chi_i>=0,
\end{equation}
that can be rewritten in the form
\begin{equation}\label{reducedCCE}
< \chi_f| ({\bf h} \cdot \hat{{\bf j}}({\bf h}))| \chi_i>=
\frac{M_f-M_i}{ M_i+M_f }< \chi_f| {\hat{\bf j}}^0({\bf h})|
\chi_i>,
\end{equation}
as far as ${\bf G}_i=-{\bf h}G_{i\,0}$, ${\bf G}_f={\bf h}G_{i\,0}$,
${\bf P}_i=M_i {\bf G}_i$, ${\bf P}_f=M_f {\bf G}_f$, $\hat{M} |
\chi_i>=M_i | \chi_i>$, $\hat{M} | \chi_f>=M_f | \chi_f>$. The
current (\ref{exact_current}) does not satisfy
Eq.~(\ref{reducedCCE}) and needs a modification. Following
\cite{Lev} we use the unique decomposition into longitudinal and
transverse parts:
\begin{equation}
{\hat{\bf j}}({\bf h})= {\hat{\bf j}}(0)+ \frac{{\bf h}}{h}
{\hat{j}}_{||}({\bf h})+ {\hat{\bf j}}_{\perp}({\bf h}),
\end{equation}
where ${\bf h}{\bf j}_{\perp}({\bf h})=0$ and
\begin{eqnarray}
&&{\hat j}_{||}({\bf h})=\frac{1}{|{\bf h}|}
\left({\bf h}\cdot({\hat {\bf j}}({\bf h})-{\hat {\bf j}}(0))\right),\nonumber\\
&&{\hat {\bf j}}_{\bot}({\bf h})={\hat {\bf j}}({\bf h})- {\hat {\bf
j}}(0)-\frac{{\bf h}}{|{\bf h}|^2}\left({\bf h}\cdot({\hat {\bf
j}}({\bf h})-{\hat {\bf j}}(0))\right). \label{decomposition}
\end{eqnarray}
To estimate violation of the current conservation equation we assume
that $NN$ interaction does not change transverse and time components
of operator  ${\hat{ j}}({\bf h})$. Then we can reconstruct  ${\hat{
\bf j}}(0)$ and  ${\hat{ j}}_{||}({\bf h})$ from
Eq.~(\ref{reducedCCE}). In the transverse gauge (\ref{trgauge}) the
longitudinal component has no effect on results of our calculation
and therefore we determine only matrix element of ${\hat{\bf j}}(0)$
\begin{equation}\label{reducedCCE}
< \chi_f| {\hat{\bf j}}(0)| \chi_i>= \frac{M_f-M_i}{ M_i+M_f }<
\chi_f| \left.\frac{\partial{\hat{\bf j}}^0({\bf h})}{\partial{\bf
h}}\right|_{h=0}| \chi_i>,
\end{equation}
 Corresponding addend $\delta {\bf j}$ that restores
Eq.~(\ref{reducedCCE}) is given in Eq.~(\ref{dj}).

The first term in Eq. (\ref{matixElAll}) (PlWA) appears as
\begin{widetext}
\begin{eqnarray}
\label{eq_17}
\langle\phi_f\vert\hat{j}^{\mu}\vert\chi_i\rangle_{n.r.}=\sqrt{\frac{2}{\pi}}\frac{1}{q_{f}}\sum_{J=0}^{3}
\sum_{l=0,2}\sum_{m=-l}^{l}i^{l} {\cal C}^{JM_{J}}_{lm\,1\mu}
\nonumber
\left(\sum_{k=1}^{4}\sum_{i=1}^{2}{\cal
Y}^{\ast}_{lm}(\hat{q}_{i})\langle l,S;JM_{J}\vert{
L}_{ki}^{\mu}\vert l,1;1M_{J} \rangle
U_l^i \right.\nonumber\\
+{\cal
Y}^{\ast}_{lm}(\hat{q}_{f})\left(1-\delta_{0,\mu}\right)\left(\sum_{i=1}^{3}\langle
l,S;JM_{J}\vert{ K}_{i}^{\mu}\vert l,1;1M_{J} \rangle
U_l(q_f)\right.\nonumber\qquad\\
\left.\left. -2g^{pn}_{e}\,w(q_{f})\,\sum_{l'=1,3}\langle
l',1;JM_{J}\vert\hat{r}^{\mu}\vert l,1;1M_{J} \rangle
\int_{0}^{\infty}r\hat{j}_{l'}(q_{f}r)u_{l}(r)dr\right)\right)
,\nonumber\\
U_l(q_f)=\int_{0}^{\infty}\hat{j}_{l}(q_{f}r)u_{l}(r)dr,\ \
\nonumber\\ U_l^i=\frac{w(q_f)(1-h^2)}{w(q_f)(1+h^2)+(-1)^i 2({\bf
h}\cdot {\bf q}_f)}\int_{0}^{\infty}\hat{j}_{l}({d}_i({\bf
q}_{f})r)u_{l}(r)dr,\qquad
\end{eqnarray}
\end{widetext}%
here ${ L}_{ki}^{\mu}\,(k=1,2,3,4)=B_{1i}^{\mu}$, $({\bf
B}_{2i}^{\mu}{\bf  s}_2)$, $({\bf  B}_{3i}^{\mu}{\bf  s}_1)$ and
$({\bf  B}_{4i}^{\mu}{\bf  s}_2)({\bf  B}_{5i}^{\mu}{\bf  s}_1)$,
respectively; $K_i^{\mu}$ represent the $\mu$-components
($\mu=1,2,3$) of the the first three ($i=1,2,3$) terms in Eq.
(\ref{dj}).

The second term in Eq. (\ref{matixElAll}) appears as
\begin{widetext}
\begin{eqnarray}
\label{eq_16} \langle \chi_f-\phi_f\vert\,\hat{\tilde{j}}^{\mu}({\bf
h})\,\vert \chi_i\rangle_{n.r.}= \sqrt{\frac{2}{\pi}}
\frac{1}{q_{f}}\sum_{L=0,2}\sum_{J=0}^{\infty}\sum_{l=J-S}^{J+S}\sum_{l'=J-S}^{J+S}\sum_{m=-l}^{l}i^{l'}
{\cal C}^{JM}_{lm\,S\mu}{\cal Y}^{\ast}_{lm}(\hat{q}_{f})\times\nonumber\\
\times\int_{0}^{\infty} dr\langle l',S;JM
\vert(u^{J}_{l',l}(q_{f},r)-\delta_{l,l'}\hat{j}_{l}(q_{f}r))\hat{\tilde{j}}^{\mu}({\bf
h})  u_{L}(r)\vert L,1;1M_{i} \rangle\,
\end{eqnarray}
\end{widetext}

We use further the algebraic results (A24)-(A27) of Ref.
\cite{Myppg2} and obtain the final expression for the differential
cross-section which is reduced to radial integrals and spherical
harmonics but, unfortunately, is too much unwieldy to be exposed
here.

Now, by a few examples, we illustrate the calculation technique for
the matrix elements of various components of the relativistic
current operator%
\begin{eqnarray}
\langle l_f,S_f;J_{f}M_{f}\vert({\bf B}^{\mu}_{3i}\cdot{\bf s}_1
)\vert l_i,S_{i};J_{i}M_{i} \rangle=\nonumber\\=({\bf
B}^{\mu}_{3i}\cdot\langle l_f,S_f;JM_{f}\vert{\bf s}_1 \vert
l_i,S_{i};J_{i}M_{i} \rangle),\nonumber
\end{eqnarray}
\begin{eqnarray}
\langle l_f,S_f;JM_{f}\vert({s}_1)_{\nu} \vert l_i,S_{i};J_{i}M_{i}
\rangle=\nonumber\qquad\qquad\qquad\\=(-1)^{L_i+J_f+S_f+1}\delta_{L_i
L_f} {\cal C}^{J_{i}M_{i}}_{1\nu\,J_{f}M_{f}}
\sqrt{2J_i+1}\times\nonumber\\\times \left\{\begin{array}{ccc}
L_{f}&J_{f}&S_{f}\\
1&S_{i}&J_{i}
\end{array}\right\}\langle S_f\vert\vert s_1 \vert\vert S_i\rangle,\nonumber\\
\langle S_f\vert\vert s_1 \vert\vert
S_i\rangle=(-1)^{S_{f}}\sqrt{(2S_i+1)(6S_f+3)/2}\times\nonumber\\\times\left\{\begin{array}{ccc}
1/2&1/2&S_{f}\\
1&S_{i}&1/2
\end{array}\right\}.\qquad
\end{eqnarray}
\begin{eqnarray}
({\bf a}_1 \cdot {\bf s}_1)({\bf a}_2 \cdot {\bf s}_2)=\qquad\qquad
\qquad\qquad \nonumber\\ \sum_{k=0}^{2} C_{k} \left[\left[a_1\times
a_2\right]^{(k)} \times\left[{s}_1\times
{s}_2\right]^{(k)}\right]^{(0)}, \nonumber
\end{eqnarray}
\begin{equation}
C_{k}=(1,\,\sqrt{3},\, \sqrt{5});\qquad
\end{equation}
\begin{eqnarray}
\langle S_f\vert\vert [s_1 \times s_2]^{(k)}\vert\vert
S_i\rangle=\nonumber\qquad\qquad\\\frac{3}{2}\sqrt{(2S_i+1)(2S_f+1)(2k+1)}\left\{\begin{array}{ccc}
1/2&1/2&S_{f}\\
1/2&1/2&S_{i}\\
1&1&k
\end{array}\right\}\nonumber.
\end{eqnarray}
\begin{eqnarray}
({\bf  \nabla} \cdot {\bf S}) \nabla_\mu =- \frac{1}{ \sqrt{3}}
\left[ [ { \nabla} \times { \nabla}]^{(0)} \times {S}
\right]^{(1)}_\mu -  \nonumber\\-\sqrt{\frac{5}{3}} \left[ [ {
\nabla} \times { \nabla}]^{(2)} \times {S} \right]^{(1)}_\mu.
\end{eqnarray}
\begin{eqnarray}
 ( {\bf h} \cdot [{\bf  \nabla} \times {\bf S}]) \nabla_\mu =\qquad\qquad\qquad\qquad\nonumber\\ -i
\frac{ \sqrt{6}}{3} \left( \frac{ \sqrt{15}}{2} \left[ [[ { \nabla}
\times { \nabla}]^{(2)} \times {S}]^{(2)} \times {h}
\right]^{(1)}_\mu \right.\nonumber\\%
+ \left[ [[ { \nabla} \times { \nabla}]^{(0)} \times {S}]^{(1)}
\times {h} \right]^{(1)}_\mu\nonumber\qquad\qquad\\
\left. - \frac{ \sqrt{5}}{2}\left[ [[ { \nabla} \times{
\nabla}]^{(2)} \times {S}]^{(1)} \times {h} \right]^{(1)}_\mu
\right).\qquad\qquad
 \end{eqnarray}
 \begin{widetext}
$$
<L_f,S_f=1;J_fM_f \mid \left[ [ { \nabla} \times { \nabla}]^{(k)}
\times {S}]^{(1)} \right]^{(n)}_\mu f(r) \mid L_i,S_i=1;J_iM_i>=
$$
\begin{equation}
C^{J_fM_f}_{J_iM_in \mu} \left\{
\begin{array}{ccc}
L_f&1&J_f\\
L_i&1&J_i\\
k&1&n
\end{array}
 \right\}
\sqrt{6(2J_i+1)(2n+1)} <L_f \| [ { \nabla} \times {
\nabla}]^{(k)}f(r)  \|L_i>,
\end{equation}
$$
<L_f \| [ { \nabla} \times { \nabla}]^{(2)} \frac{f(r)}{r} \|L_i>=
$$
$$
\frac{ \sqrt{2L_f+1}}{ \sqrt{6}C^{L_f0}_{L_i020}} \frac{1}{r} \left(
\delta_{L_iL_f} \left( -1+ \frac{3(2L_i^2+2L_i-1)
\sqrt{2(2L_i+1)}}{(2L_i-1)(2L_i+1)(2L_i+3)} \right) \left(
\frac{d^2}{dr^2}- \frac{L_i(L_i+1)}{r^2}  \right)f(r) \right.
$$
$$
+ \delta_{L_iL_f-2} \frac{3(L_i+1)(L_i+2)
\sqrt{2(2L_i+1)}}{(2L_i+1)(2L_i+3)(2L_i+5)} \left( \frac{d^2}{dr^2}-
\frac{(2L_i+3)}{r} \frac{d}{dr}+ \frac{(L_i+3)(L_i+1)}{r^2}
\right)f(r)
$$
\begin{equation}
\left. + \delta_{L_iL_f+2} \frac{3L_i(L_i-1)
\sqrt{2(2L_i+1)}}{(2L_i+1)(2L_i-3)(2L_i-1)} \left( \frac{d^2}{dr^2}-
\frac{(2L_i-1)}{r} \frac{d}{dr}+ \frac{L_i(L_i-2)}{r^2} \right) f(r)
\right) .
\end{equation}
\end{widetext}
In these expressions, an upper index in round brackets means a
tensor rank of an operator. The first rank is omitted where it is
obvious  ($\nabla\equiv\nabla^{(1)}$ \textit{etc.})

\section{\label{sec:APPENDIXB} POINT-FORM MOMENTUM TRANSFER}
In the general case there are initial $NN$-state with associated
initial c.m. frame (i.c.m.f.) and final $NN$-state  with associated
final c.m. frame (f.c.m.f.) Suppose that the photon momentum
(momentum transfer) is along the $z$ axis. Values of photon momentum
and energy in i.c.m.f. are $|{\bf q}_{\gamma}|$ and $q_{\gamma}^0$
correspondingly. Momentum transfer is $Q^2=|{\bf q}_{\gamma}|^2-
(q_{\gamma}^0)^2$. Let $P$ be the total 4-momentum of the
$NN$-system, $M$ be the mass of the $NN$-system, $G=P/M$ be the
system 4-velocity. Index $i(f)$ means initial (final) state of the
$NN$-system. Transformation from i.c.m.f. to the special frame
suggested by Lev \cite{Lev} (L.s.) where
\begin{equation}
{\bf G}_f+{\bf G}_i=0|_{L.s.}\label{Levscm}%
 \end{equation}
is defined by angle $\Delta/2$ such that
\begin{equation}
\tanh\Delta/2=h, \label{Deltafromh}
\end{equation}
where ${\bf h}={\bf G}_f/G^{0}_f|_{L.s.}$. The Lev frame
(\ref{Levscm}) is not equivalent to the Breit frame  defined by the
condition ${\bf P}_f+{\bf P}_i=0$ if $M_f\neq M_f$. In case of
elastic electron-deuteron scattering these frames coincide.

From this point we may use a special derivation of Ref. \cite{Allen}
(Eqs.~(\ref{initialmomenta}-\ref{transferMy}) of the present paper).

The initial energies and z-components of momenta in L.s. are
\begin{eqnarray}
w_1=w\cosh \Delta/2+q_{z}\sinh\Delta/2\nonumber\\
q_{1z}=q_{z}\cosh \Delta/2+w\sinh\Delta/2\nonumber\\
w_2=w\cosh \Delta/2-q_{z}\sinh\Delta/2\nonumber\\
q_{2z}=-q\cosh \Delta/2+w\sinh\Delta/2,\label{initialmomenta}
\end{eqnarray}
where ${\bf q}$ and $w=\sqrt{{\bf q}^2+m^2}$ are center of momentum
variables, ${\bf q}$ is momentum of particle one (internal
variable). After the photon absorption the $z$-component of the
internal variable and corresponding energy change
\begin{eqnarray}
q_{z}'=q_{z}\cosh \Delta\mp w\sinh\Delta\\
w'=w\cosh \Delta\mp q_{z}\sinh\Delta,
\end{eqnarray}
where the minus (plus) sign is used when particle one (two) is
struck. The final energies and momenta in L.s. will then be
\begin{eqnarray}
w_1'=w\cosh 3\Delta/2-q_{z}\sinh 3\Delta/2\nonumber\\
q_{1z}'=q_{z}3\cosh \Delta/2-w\sinh3\Delta/2\nonumber\\
w_2'=w_2;\ \ \ \ \ q_{2z}'=q_{2z}, \label{new_moment}
\end{eqnarray}
other components do not change. Some hyperbolic trigonometry reveals
that
\begin{equation}
(q_1'-q_1)^2=4(q_{z}^2-w^2)\sinh^2\Delta\label{transferMy},%
 \end{equation}
it follows from Eq.~(\ref{Deltafromh}) that
\begin{eqnarray}
\sinh\Delta=\frac{2h}{1-h^2}.%
 \end{eqnarray}
Since
\begin{equation}
q_{z}^2-w^2=-(m^2+{\bf q}^2_{\perp})=-(m^2+{\bf q}^2-\frac{({\bf q}\cdot{\bf h})^2}{h^2}),\label{q_zaction}%
 \end{equation}
 the resulting Eq.~(\ref{fatransfer}) is established.
%

\end{document}